\documentclass{aa}  
\usepackage{graphicx}
\usepackage{txfonts}
\usepackage{hyperref}
\hypersetup{
    colorlinks=true,
    linkcolor=blue,
    filecolor=magenta,      
    urlcolor=cyan,
    pdftitle={Overleaf Example},
    pdfpagemode=FullScreen,
    linkcolor=blue,  
    citecolor=blue,        
}
\usepackage{bm}
\usepackage{lipsum} 
\usepackage[caption=false]{subfig}
\usepackage[table,xcdraw]{xcolor}
\usepackage{booktabs}
\usepackage{multirow}
\usepackage{bbm}
\usepackage[normalem]{ulem}

\let\orgautoref\autoref
\providecommand{\Autoref}[1]{\def\equationautorefname{Equation}
\def\figureautorefname{Figure}\def\sectionautorefname{Section}\def\subsectionautorefname{Section}\def\subsubsectionautorefname{Section}\def\tableautorefname{Table}\orgautoref{#1}}
\renewcommand{\autoref}[1]{\def\equationautorefname{Eq.}
\def\figureautorefname{Fig.}\def\sectionautorefname{Sect.}\def\subsectionautorefname{Sect.}\def\subsubsectionautorefname{Sect.}\def\tableautorefname{Table}\orgautoref{#1}}

\begin{document} 

    \title{Breaking the degeneracy in stellar spectral classification from single wide-band images}


    \author{
        Ezequiel Centofanti \inst{1},
        Samuel Farrens \inst{1},
        Jean-Luc Starck\inst{1,2},
        Tobias Liaudat\inst{3},
        Alex Szapiro \inst{1,4},
        Jennifer Pollack \inst{1}
        }

    \institute{
             Université Paris-Saclay, Université Paris Cité, CEA, CNRS, AIM, 91191, Gif-sur-Yvette, France
             \and
             Institutes of Computer Science and Astrophysics, Foundation for Research and Technology Hellas (FORTH), Greece 
             \and
             IRFU, CEA, Université Paris-Saclay, F-91191 Gif-sur-Yvette, France
             \and
             IMT Atlantique, École Mines-Télécom, Bretagne-Pays de la Loire, France
             \\
    \vspace{0.001cm}\\
    \email{ezequiel.centofanti@cea.fr}
    }

\abstract{ 

    The spectral energy distribution (SED) of observed stars in wide-field images is crucial for chromatic point spread function (PSF) modelling methods, which use unresolved stars as integrated spectral samples of the PSF across the field of view. 
    This is particularly important for weak gravitational lensing studies, where precise PSF modelling is essential to get accurate shear measurements. 
    Previous research has demonstrated that the SED of stars can be inferred from low-resolution observations using machine-learning classification algorithms. However, a degeneracy exists between the PSF size, which can vary significantly across the field of view, and the spectral type of stars, leading to strong limitations of such methods.
    We propose a new SED classification method that incorporates stellar spectral information by using a preliminary PSF model, thereby breaking this degeneracy and enhancing the classification accuracy. Our method involves calculating a set of similarity features between an observed star and a preliminary PSF model at different wavelengths and applying a support vector machine to these similarity features to classify the observed star into a specific stellar class. The proposed approach achieves a 91\% top-two accuracy, surpassing machine-learning methods that do not consider the spectral variation of the PSF. 
    Additionally, we examined the impact of PSF modelling errors on the spectral classification accuracy.
    
    }

\keywords{ Stellar classification -- Point spread function -- Machine learning}
    
\titlerunning{PSF-aware spectral classification}
\authorrunning{E. Centofanti}
\maketitle
  
    \section{Introduction}
    \label{sect:intro}

    Significant efforts have been made in recent decades to study the stars in our galaxy in detail.
    For this purpose, several spectroscopic surveys have been carried out, which provide very high-resolution catalogues of stellar spectra \citep{UVES-POP,ELODIE,Pickles_1998,MILES,INDO-US_CFLIB}. Stars are generally categorised according to the Morgan-Keenan (MK) system \citep{1943assw.book.....M}. 
    This system segments stars in descending order of temperature, with O-type stars being the hottest and M-type stars the coolest. Each of these stellar classes has a distinctive spectrum, the main characteristics of which are shared between stars of the same class. 
    Stellar classification is crucial for studying the composition of stars and their properties, and for understanding the evolution of the galactic stellar population.
    One particularly interesting case where stellar spectra play a significant role is in modelling the instrumental response of wide-field, single-band telescopes.
    
    Current space telescopes such as \textit{Euclid} \citep{euclidcollaboration2024euclidiovervieweuclid, laureijs} or upcoming telescopes such as the Nancy Grace \textit{Roman} space telescope \citep{akeson2019widefieldinfraredsurvey, spergel2015widefieldinfrarredsurveytelescopeastrophysics} and the Vera C. Rubin Observatory \citep{Ivezi__2019, lsstsciencecollaboration2009lsstsciencebookversion} will observe the Universe with unprecedented accuracy and coverage, providing vast amounts of data in the coming decades that will drive cosmology forward and enable new discoveries to be made.  
    Given their depth and coverage, these surveys will be able to examine the large-scale structure of the late-time Universe using statistical probes, such as Weak gravitational Lensing (WL).
    The WL signal is measured by correlating the shape and orientation of large numbers of galaxies. The apparent shape of the observed galaxies is corrupted by the instrumental response of the telescope, which directly limits the quality of the measured WL signal \citep{10.1093/mnras/sts371}.
    Hence, an accurate model of the point spread function (PSF), that is, the instrumental response of the optical system, is a fundamental requirement for obtaining unbiased and competitive constraints on cosmological parameters. Cutting-edge space telescopes have such low aberrations that the instrumental response is diffraction limited and is mainly driven by the optical system \citep{euclidcollaboration2024euclidiivisinstrument}. 
    Even so, in addition to having a spatial and temporal dependence, the PSF is also strongly wavelength-dependent, which is particularly challenging for wide single-passband instruments.
    Most PSF modelling methods \citep[for a review see e.g.][]{Liaudat_2023}, in particular data-driven methods, such as \texttt{PSFEx} \citep{2011ASPC..442..435B}, RCA \citep{Ngole_2016}, MCCD \citep{refId0}, and WaveDiff \citep{liaudat}, make use of observations of unresolved stars to sample the underlying instrumental response of the telescope.
    These samples allow the PSF model to be constrained at various positions in the field of view (FOV).
    Chromatic PSF models \citep[e.g.][]{liaudat} additionally require knowledge of the SED of the stars that are used to fit the model to account for the spectral dependence of the PSF of \textit{Euclid}-like (i.e. optical single-band wide-field) telescopes.
    
    Given the temporal variation caused by temperature fluctuations and the resulting mechanical stress on the optical system of space telescopes, the PSF model has to be recalibrated for each individual exposure using only the unresolved stars present in that exposure.
    Additionally, stellar spectra are typically measured for a small number of stars (at most several hundred), which is largely insufficient for chromatic PSF modelling. Astrometric missions, such as GAIA \citep{ 2016A&A...595A...1G, 2001A&A...369..339P}, can provide spectral information for a considerably larger number of stars, albeit at a lower spectral resolution. 
    Nevertheless, these complementary surveys will only measure the brightest stars in the FOV of a \textit{Euclid}-like exposure. Therefore, the availability of spectral information for the observed stars is a limiting factor in the estimation of data-driven chromatic PSF models.
    
    Stellar classification from photometric observations provides a way to increase the number of available stars for PSF modelling. There are numerous methods to classify stellar spectra with spectroscopic data, such as in \citet{10.1093/mnras/stz3100} and \citet{2023MNRAS.518.3123C}. Other studies, such as \citet{2024Univ...10..214Y}, propose classification from multi-band photometric observations. 
    Previous work by \citet{kuntzer} introduced a stellar classifier that assigns spectral templates to single-band star observations. Their method is based on a principal component analysis (PCA) decomposition of the observed stars followed by a fully connected neural network classifier. 
    Each star to be classified is projected onto the PCA feature space and the associated coefficients are fed into a multi-layer perceptron (MLP) network \citep{bishop1995neural}. The MLP classifies the coefficients into a given stellar class, assigning a specific spectral template to the observation. Their work demonstrated that it is possible to perform stellar classification with single wide-band star images. However, the methodology they propose does not account for the degeneracy between the size of the PSF and the spectral type of the star. This degeneracy (detailed in \autoref{sect:psf_aware}) refers to the fact that stars at different positions in the sky with different stellar types can have very similar image properties (i.e. size and shape) due to variations in the PSF across the field of view. As the classification method only has access to the image properties, there is an inherent limitation in the accuracy that can be achieved.

    In this work, we consider stellar classification from single-band photometric images for a \textit{Euclid}-like survey.
    We propose a stellar spectral classification method from single-band star observations that takes into account the PSF spectral variation by using a preliminary PSF model, thus breaking the degeneracy between the size of the PSF and the spectral type of the star. We only consider the WaveDiff PSF model since it is the sole data-driven model that accounts for the chromatic variation of the PSF. Firstly, we apply the PCA+MLP classification method introduced by \citet{kuntzer} to \textit{Euclid}-like simulated stellar images to establish a baseline for comparison. Secondly, we introduce a modified version of their method, where the PCA decomposition is replaced by a convolutional neural network (CNN) that takes the single-band stellar image pixels and outputs a feature vector, which is then similarly fed into an MLP network that assigns the stellar type. Finally, we introduce our new PSF-aware classification method and compare it to the previous solutions.
    The structure of this paper is the following: in \autoref{sect:stars_in_the_fov} we describe stellar observations in the context of PSF modelling, in \autoref{sect:pixel_only_classif} we introduce the pixel-only stellar classification algorithms, in \autoref{sect:psf_aware} we present the PSF-aware classification method, in \autoref{sect:sims} we explain the simulated star observations and the PSF modelling details, in \autoref{sect:results} we show the results of the classification methods and finally, \autoref{sect:conclusion} summarises this work and outlines future steps that could be taken.
    \paragraph{Notation:}In this paper, we adopt the notation for PSF modelling and astronomical imaging defined in \citet{Liaudat_2023}. A summary of the chosen notation can be found in \autoref{tab:notation}.
    \paragraph{Terminology:}In this article, the terms polychromatic PSF or PSF sample are equivalent to a star observation (or simulation) since they include the spectral information of the star. The term
    PSF by itself refers to the underlying instrumental response of the instrument or a simulated PSF model, a function of the FOV coordinates and the wavelength.
    A monochromatic PSF is the PSF model evaluated at a single wavelength for a specific FOV position. 
    The wavefront error (WFE) is the phase difference between the incoming light wavefront and an ideal hemispherical wavefront at the pupil plane of the optical system. 

    \section{Stellar image model}
    \label{sect:stars_in_the_fov}

    PSF modelling for \textit{Euclid}-like telescopes involves several key challenges. First, the observations are integrated over the passband of the telescope (i.e. single-band or polychromatic observations), thus blending the spectral variation of the PSF with the SED of the star. Second, the observation is subsampled on the detector \citep{ 2007PASP..119.1295H, 1999PASP..111.1434L} and contains observational noise that encompasses thermal noise, readout noise \citep{10.1007/1-4020-2527-0_82}, and dark-current shot noise \citep{2006SPIE.6068...37B}. 
    Finally, the number of stars for which the SED is known is very limited \citep{10.1093/mnras/staa1818}. Thus, PSF modelling from low-resolution star observations proves to be a challenging task that would benefit from an increase in spectral information (SEDs) of distant stars in the FOV.
    
    Distant stars can be considered as point sources whose intensity varies with wavelength according to the SED of each star. 
    The observational model of a distant star at the position $(u_i, v_i)$ in the FOV is as follows
    \begin{multline}
        I_{\text{star}}(\bar{u},\bar{v}|u_i,v_i) = \\
        \mathcal{F}_p \left\{ \int_0^{+\infty} \mathcal{T}(\lambda) \text{SED}(\lambda) \; \mathcal{H}_{\text{int}}(u,v;\lambda|u_i, v_i) \; d\lambda \right\} \\
        \;+\; N(\bar{u},\bar{v}|u_i,v_i),
        \label{eq:obs_star}
    \end{multline}
    where $\mathcal{H}_{\text{int}}(u,v;\lambda|u_i, v_i)$ is the PSF of the telescope with its centre at the position of the star\footnote{In the adopted notation (\autoref{apx:notation}) the symbol $|$ means centred at the given position.}. The PSF is positively valued and has two spatial coordinates $(u,v)$ and one spectral coordinate $\lambda$.  
    The PSF sample is integrated over the passband of the telescope, given by the transmission function $\mathcal{T}(\lambda)$, together with the SED of the star. The $\mathcal{F}_p$ operator is a discretisation function that models the pixelisation of the detector (sampling) and $N$ represents the observational noise. The observed image $I_{\text{star}}$, with pixel coordinates $(\bar{u}, \bar{v})$, is a single-band discrete version of the star corrupted by the PSF of the telescope and the observational noise.
    
    Measuring stellar spectra to high precision is a challenging task that can only be carried out for a limited number of stars, generally reserved for the brightest stars in the FOV. Deep and wide space-based optical surveys can observe hundreds to thousands of unresolved stars in one exposure \citep{laureijs}. However, only a fraction of the observed stars will have complementary SED information from spectroscopic surveys.
    Thus, estimating the spectra of the remaining stars is crucial for improving PSF modelling and maximising the scientific returns of these observations.

    \Autoref{eq:obs_star} describes the observational model of an unresolved star. The PSF of a telescope at the position of the star in the FOV is integrated alongside the SED of the given star. Thus, the observation is directly affected by the spectrum of the star. O-type stars, being hotter and bluer, have higher flux at shorter wavelengths. Conversely, M-type stars, being cooler and redder, have a higher flux at longer wavelengths.  The fact that the PSF has a chromatic variation allows some of the spectral information of the stars to permeate into the observations, meaning that different types of stars at the same position in the FOV will produce different observations. Hence, a spectral classification of stars from their polychromatic (wavelength integrated or single-band) observations is physically supported. 

    \begin{figure}
        \subfloat[]{
        \includegraphics[clip,width=.99\columnwidth]{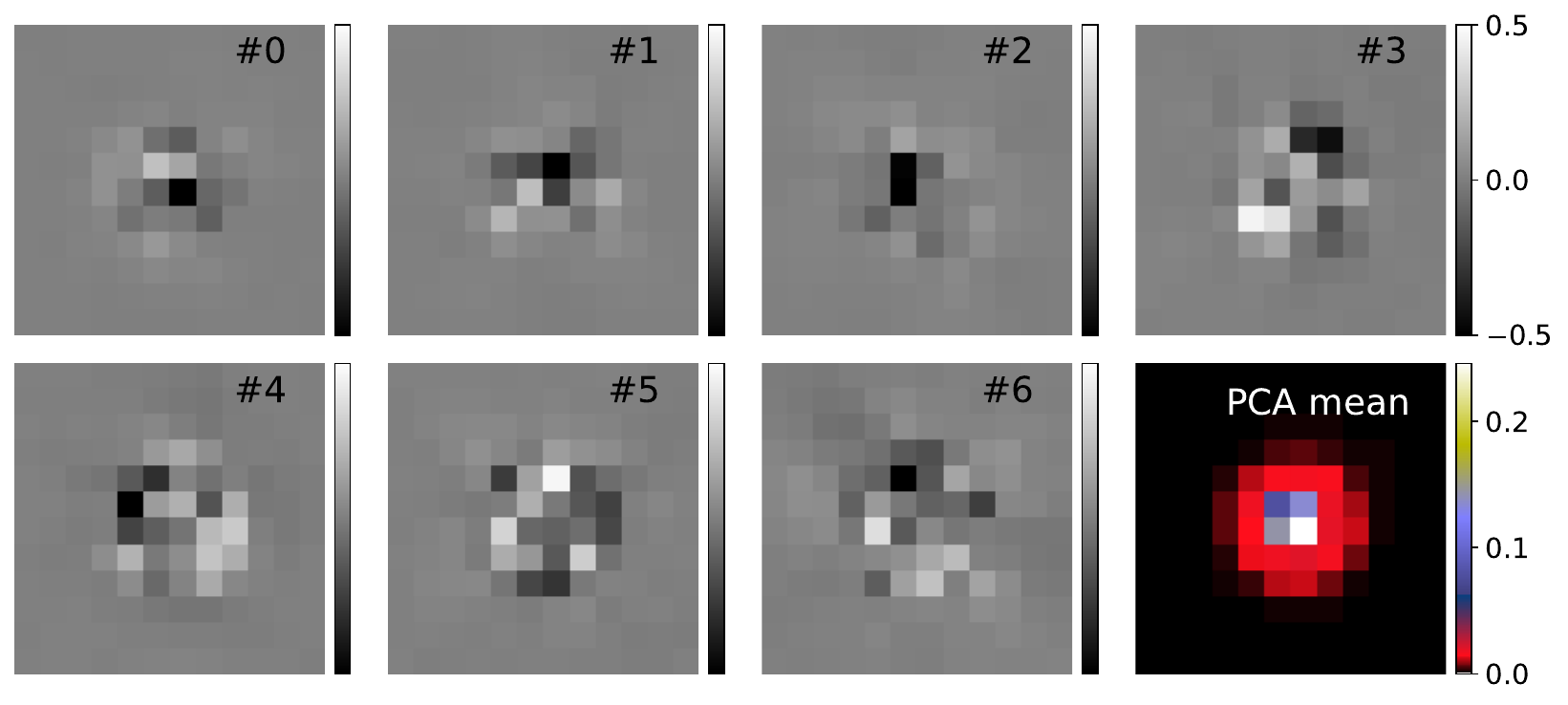}}
        
        \subfloat[]{   
        \includegraphics[clip,width=.94\columnwidth]{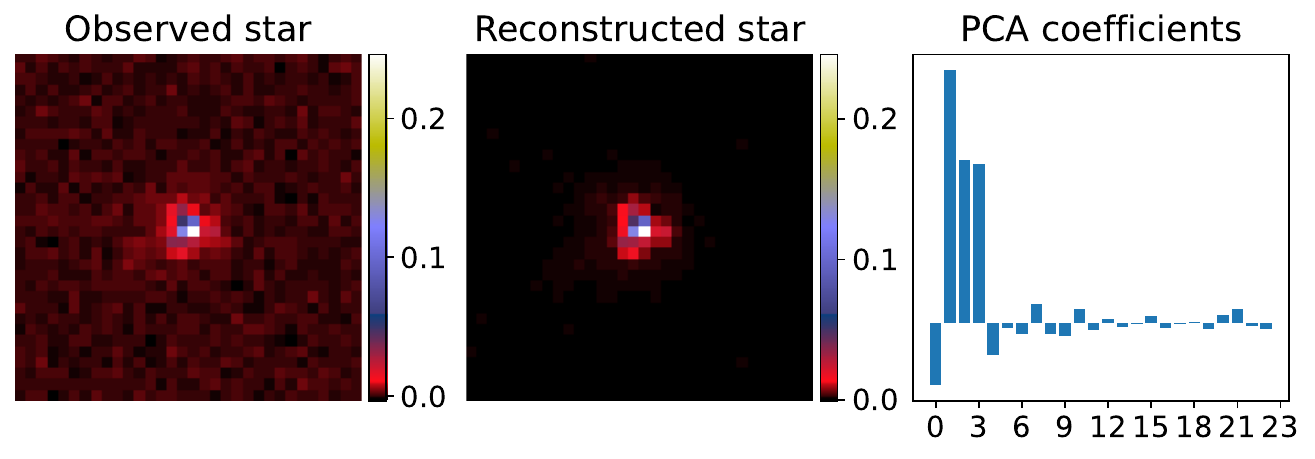}}
        \caption{PCA decomposition of input stars. (a) First seven PCA components and PCA mean. (b) Original star observation and its reconstruction from the first 24 PCA components. The last figure in the lower panel shows the relative values of the coefficients associated with the 24 PCA components.}
        \label{fig:PCA}
    \end{figure}
    
    \section{Star classification from a single wide band}
    \label{sect:pixel_only_classif}
    In the following, we present some methods of stellar classification from single wide-band star observations. The objective of these classification methods is to assign SED templates to the stars without spectral information using the pixels in the corresponding postage stamps. For all of the methods, we assume a scenario in which, given an exposure of a \textit{Euclid}-like telescope, source detection is performed with a tool, such as SExtractor \citep{1996A&AS..117..393B}, SFIND \citep{Hopkins_2002} or IMSAD \citep{1995ASPC...77..433S}. Then, these sources are efficiently classified as stars or galaxies, and postage stamps are extracted at the positions of the stars. For the purpose of this work, sources of contamination in the star selection such as galaxies, binaries, or other misclassifications are not considered. We expect only a small fraction of the detected stars to have spectral information (SEDs) available from complementary measurements. 
    
    \subsection{PCA MLP}
    The method proposed by \citet{kuntzer} can be separated into two steps: the preprocessing of the input data and the actual spectral classification. The first step aims to extract relevant structure from the observations by compressing the input into a reduced number of coefficients. This is done by applying PCA to project the input onto $24$ orthogonal components (i.e. the PCA coefficients). 
    The second step implements a fully connected MLP neural network \citep{bishop1995neural} classifier that takes as input the $24$ PCA coefficients associated with the star image to be classified, and outputs the predicted spectral class among the 13 spectral classes considered \citep{Pickles_1998}. We implemented this method from scratch\footnote{
    The code is available here:
    
    \hspace{0.36cm}\href{https://github.com/CentofantiEze/sed_spectral_classification}{https://github.com/CentofantiEze/sed\_spectral\_classification}} and adapted it to our synthetic star observations (see \autoref{sect:sims}). We use $10\,000$ simulated stars to obtain the PCA components and train the MLP classifier. We present the first seven PCA components as well as the dataset mean in the top panel (a) of \autoref{fig:PCA}. In the bottom panel (b), we show a star observation and its reconstruction from the first 24 PCA components. The relative values of the 24 coefficients are shown in the last figure in the lower panel. The reconstruction maintains the overall shape of the observed star filtering out the high frequency variations. In other words, the observation is denoised. This process significantly reduces the size of the data, compressing the $32\times32$ px images ($1\,024$ pixels) into 24 coefficients.
    As in \citet{kuntzer}, we train a committee of $48$ networks from which we compute the ensemble average of the predictions, allowing us to obtain a more robust classification. Each MLP classifier has two hidden layers of $26$ nodes each. 

    \begin{figure}
        \centering
        \includegraphics[height=.12\textwidth, trim={.8cm 0 0.7cm 0},clip]{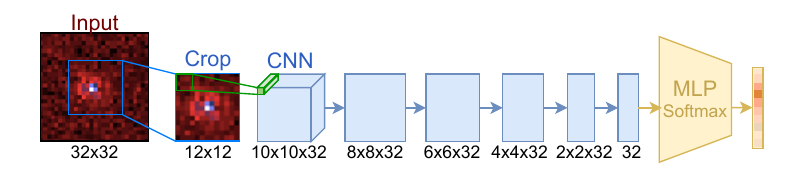}
        \caption{CNN+MLP model diagram. In light blue the convolutional blocks. In yellow the multi-layer perceptron classifier.}
        \label{fig:CNN-architecture}
    \end{figure}

    \subsection{CNN MLP}
    Convolutional neural networks \citep{ NIPS2012_c399862d, 6795724} are widely used for recognising structure in two-dimensional data and have been widely applied to astronomical data \citep{Akhaury_2022, 2022A&A...657A..98F, 2018A&A...611A...2S, 2021A&A...651A..55S}. It has been shown that the first convolutional filters of properly trained CNNs resemble classical image processing filters \citep{6126474, zeiler2013visualizingunderstandingconvolutionalnetworks}, and are able to identify patterns and the multi-scale structure of images. Auto-encoder networks \citep{pmlr-v27-baldi12a, aic.690370209} extract relevant features of the data, encoding the input into a reduced number of values. In a similar approach, we propose to replace the PCA preprocessing step in the \citet{kuntzer} approach with an encoder-like CNN network, keeping the same two-step process: preprocessing and classification.

    \Autoref{fig:CNN-architecture} shows the proposed architecture for the classifier. The first layer centre-crops the star observations to get rid of pixels that only contain noise, reducing the size of the input. Then, the cropped image passes through six convolutional layers. Each layer has $32$ channels and gradually reduces the width and height of the data. The convolutional kernel size is $3\times3$, thus reducing the data dimension by $2$ pixels per layer. The last convolutional layer kernel size is $2\times2$ given the size of the input data in this layer. The output of the last convolutional layer is a $32$-dimensional vector that encodes the main structural features of the input. These features are fed into a multi-layer perceptron with two hidden layers of $32$ neurons each and an output layer of $13$ neurons that correspond to the $13$ stellar classes. Every layer (CNN and MLP) uses a ReLU activation function, except for the output layer which has a softmax activation function. This activation function enforces the output to be a probability vector, that is, the $i$-th element of the output vector represents the probability that the input belongs to the $i$-th stellar class.
    We compare the performance of these two pixel-only classification methods on simulated data in \autoref{sect:results}.
    
    \section{Breaking the degeneracy}
    \label{sect:psf_aware}
    
    \begin{figure*}
        \centering
        \hspace*{-.2cm}
        \subfloat[]{
        \includegraphics[width=.78\textwidth, trim={1cm 0 2cm 1cm},clip]{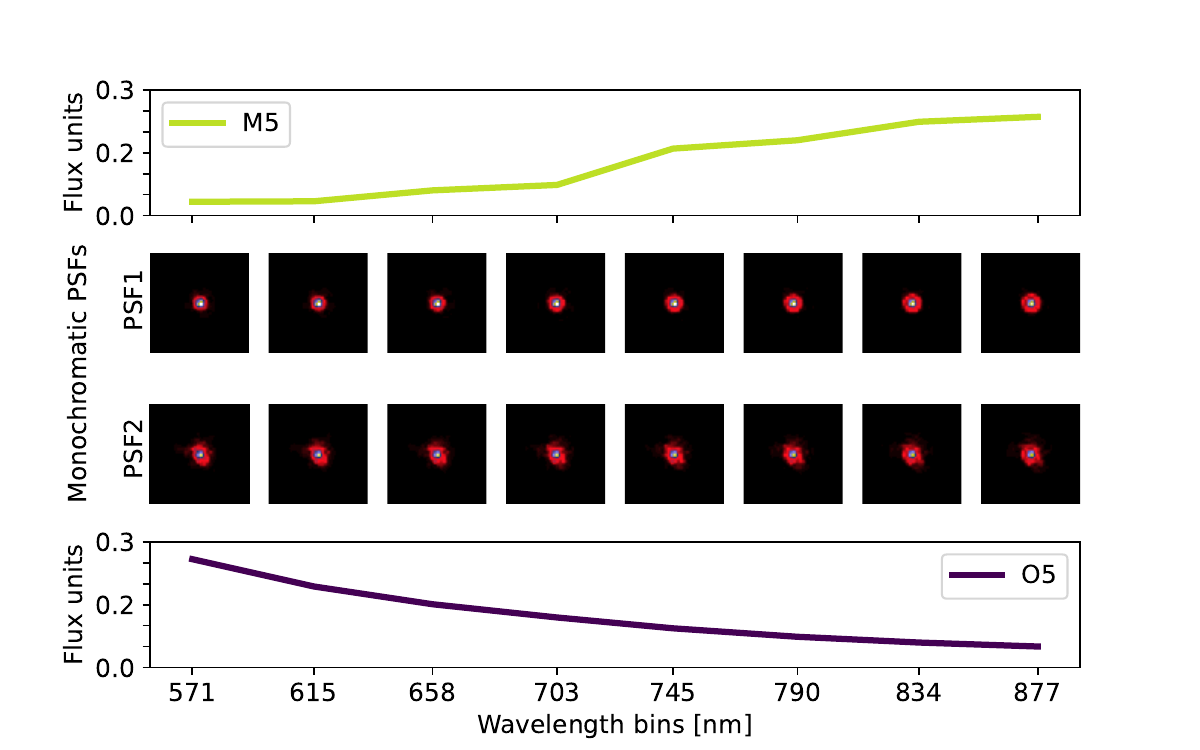}}
        \subfloat[]{   
        \includegraphics[width=.26\textwidth, trim={.5cm -1.3cm 0cm 0cm},clip]{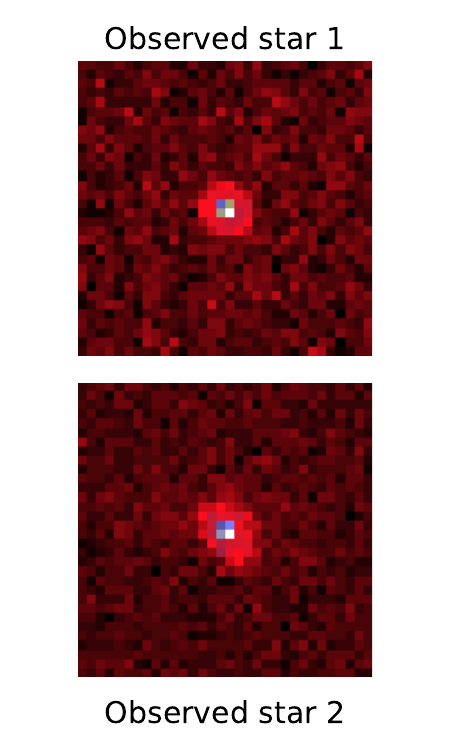}}
        \caption{Ilustrative example of the degeneracy between the PSF size and spectral type of stars. (a) Middle rows: monochromatic PSFs for two positions in the FOV. The PSFs are shown for eight equally spaced wavelength values. Top and bottom: eight-bin spectral energy distribution of two stars, a $M5$ star (top) and an $O5$ star (bottom), located at positions 1 and 2 respectively. The wavelength axis is shared between all rows. (b) Observation of the respective stars.
        }
        \label{fig:degeneracy}
    \end{figure*}
    A significant challenge for the methods described in \autoref{sect:pixel_only_classif} is the degeneracy between the PSF size and the spectral type of the star. This degeneracy is not addressed since these methods rely solely on the star image pixels for the classification, and do not include any information on the underlying PSF model corresponding to the observations. As a result, this degeneracy sets a cap on the classification accuracy of these types of methods.
    To illustrate this issue, consider the example shown in \autoref{fig:degeneracy}. The left-hand side (a) of the figure presents two different monochromatic PSFs ($\text{PSF}_1$ and $\text{PSF}_2$) at two different positions in the FOV. The chromatic variation of the PSFs is shown for eight different equally spaced wavelength values. Suppose that in each FOV position we observe an unresolved star. At FOV position 1 (top half of the figure) a $M5$-type (red) star is observed, whose eight-bin SED is shown in the top panel. At FOV position 2 (bottom half of the figure) an $O5$-type (blue) star is found, with its corresponding eight-bin SED in the bottom panel. According to a discrete version of \autoref{eq:obs_star}, where the integral is approximated by a summation, the star observations are the sum of the monochromatic PSFs weighted by the corresponding SED values (see \autoref{eq:sim_star}). 
    The fact that $\text{PSF}_1$ at long wavelengths is similar in size to $\text{PSF}_2$ at shorter wavelengths produces observations of similarly shaped stars, even if they correspond to completely different stellar types. The right-hand side of the figure (b) shows the corresponding observed stars. The observed star at position 2 has a similar, if not larger, size than star at position 1. In principle, we would associate a larger shape with a redder (M-type) star\footnote{
    It is important to note that the observed stars are unresolved and therefore appear as point sources. The apparent sizes discussed here are due to the PSF at different positions in the FOV and at different wavelengths, not the physical sizes of the stars themselves.}, which is the opposite of what is shown in this example.
    
    This example highlights that it is very difficult, if not improbable, to make a highly accurate spectral classification of stars from single-band observations alone. While many stars can be correctly classified, the confusion introduced by this degeneracy can only be overcome if the spatial and spectral variation of the PSF are considered along with the single-band observations.
    Two key issues must be considered in this regard. On the one hand, we need to know with a good level of precision the PSF (including its spatial and spectral variations) of the telescope with which the observations are obtained. On the other hand, we have to come up with an architecture capable of taking as input both the star observation and the approximate PSF model at the corresponding star position. We elaborate on these two issues in the following subsections.

    \subsection{Approximate PSF model}
    \label{sect:psf_modeling}
        
    To generate the simulated observations (see \autoref{sect:sims}) we model a ground truth (GT) PSF representing the instrumental response of the telescope. 
    This GT model is assumed to be completely unknown when processing the simulated observations (i.e. performing the classification). To obtain an approximation of this GT model we have to fit a PSF model to the simulated observations. 
    The approximate PSF model $\tilde{H}$ is estimated with a reduced number of observations since the SEDs are not available for all the observed stars prior to the stellar classification.
    In this work, we use the WaveDiff PSF modelling software \citep{liaudat}. Details about the WaveDiff PSF model can be found in \autoref{apx:wavediff}. 
    
    \begin{figure}
        \centering
        \includegraphics[width=.99\columnwidth, trim={.7cm 0 2cm 0},clip]{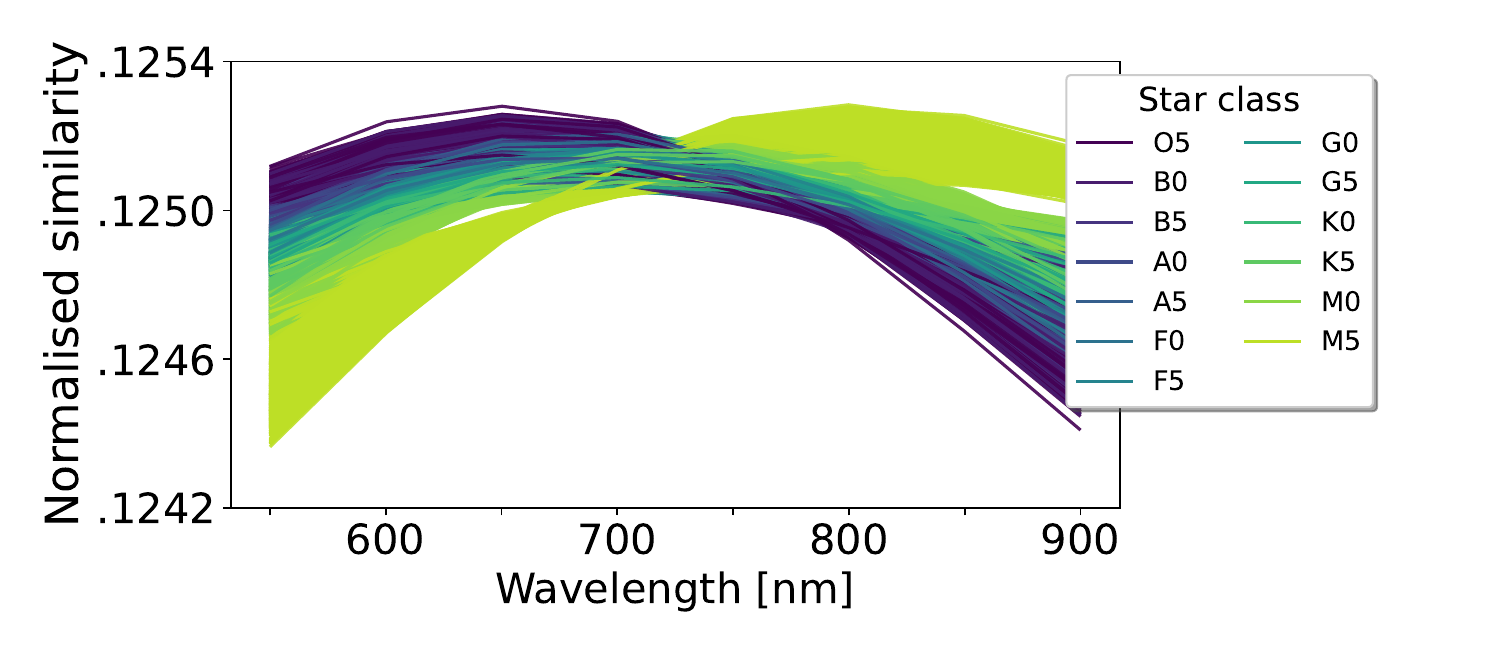}
        \caption{Normalised similarity features as a function of monochromatic PSF wavelength for the $10\,000$ stars. Each set of features is coloured according to the corresponding observation stellar type.}
        \label{fig:similarity}
    \end{figure}
    
    \subsection{Similarity features}
    \label{sec:sim_feat}

    There are many ways in which one could envision adapting the classification models already presented to take into account the PSF. Our approach is perhaps the simplest and is based on the approximate star observation model, where the integral in \autoref{eq:obs_star} is approximated by the sum of the monochromatic PSFs weighted by the SED of the star
    \begin{equation}
        I_{\text{star}}(\bar{u},\bar{v}|u_i,v_i) = \sum_{k = 1}^{n_\lambda} \text{SED}(\lambda_k)\; 
        H(\bar{u},\bar{v};\lambda_k|u_i,v_i)
        + \; N,
        \label{eq:sim_star}
    \end{equation}
    where $n_{\lambda}$ is the number of wavelength bins centred at $\lambda_k$. We can expand this summation for each bin as follows
    \begin{multline}
        I_{\text{star}}(\bar{u},\bar{v}|u_i,v_i) =\\
        b_0 \; H(\bar{u},\bar{v};\lambda_0|u_i,v_i) + 
        ... + b_n \; H(\bar{u},\bar{v};\lambda_n|u_i,v_i) + N,
    \label{eq:observation_sum}
    \end{multline}
    where $H(\bar{u},\bar{v};\lambda_k|u_i,v_i)$ are the monochromatic PSFs (i.e. evaluated at a single wavelength) at the position of the star and $b_k$ are the SED values. We use the approximate PSF model $\tilde{H}$, evaluated at the position of the star, to compute a similarity metric, referred to as similarity features (SF), for each wavelength $\lambda_k$ by comparing the approximate monochromatic PSFs with the observation,
    \begin{multline}
        \text{SF}\left\langle I_{\text{star}}(\bar{u},\bar{v}|u_i,v_i) \; ; \tilde{H}(\bar{u},\bar{v};\lambda_k|u_i,v_i) \right\rangle (\lambda_k|u_i, v_i) = \\
        \frac{
        1 -  
        \| I_{\text{star}}(\bar{u},\bar{v}|u_i,v_i) - 
        \tilde{H}(\bar{u},\bar{v};\lambda_k|u_i,v_i) \|^2_{\bar{uv}}}{
        n_\lambda - \sum_{j=1}^{n_\lambda}
        \|I_{\text{star}}(\bar{u},\bar{v}|u_i,v_i) - 
        \tilde{H}(\bar{u},\bar{v};\lambda_j|u_i,v_i) \|^2_{\bar{uv}}},
    \label{eq:similarity}
    \end{multline}
    where $\|\cdot\|^2_{\bar{uv}}$ is the Frobenius matrix squared norm over the image pixels,
    \begin{equation}
        \|I_{\text{img}}(\bar{u}, \bar{v}|u_i,v_i)\|^2_{\bar{uv}}=\sum_{\bar{u},\bar{v}=1}^{N_{pix},N_{pix}} |I_{\text{img}}(\bar{u}, \bar{v}|u_i,v_i)|^2.
        \label{eq:frobenius}
    \end{equation}
    The resulting similarity features serve as a proxy for the SED values $b_k$. Therefore, we expect higher similarity in the bins that contributed the most to the star observation, that is, to the weighted sum of monochromatic PSFs (\autoref{eq:observation_sum}). 
    
    \Autoref{fig:similarity} shows the similarity features, coloured according to the corresponding stellar class, of the full classification dataset ($10\,000$ simulated stars) using the ground truth PSF model.
    The visible distinction between the curves for each stellar type demonstrates a clear correlation between spectral type and similarity features. This strongly suggests that the similarity metric extracts relevant spectral information from the single-band observations. The dispersion in the similarity feature distribution (\nobreak\autoref{fig:similarity}) for stars of the same spectral class is driven by the spectral variation of the PSF for each star, and more specifically how alike the monochromatic PSFs are to each other at that position in the FOV.
    
    \subsection{SVM classifier}
    
    We use the similarity features as the input to our SED classifier. This greatly reduces the complexity of the classifier and thus the computational needs for training and inference. We use the C-Support Vector Classification model from the \texttt{sklearn.svm} \citep{pedregosa2018scikitlearnmachinelearningpython} Python library. This algorithm allows us to find optimal boundaries in the $n_\lambda$-dimensional similarity feature space for classifying the data into the $13$ stellar classes. If the data are linearly separable, the algorithm finds optimal separating hyperplanes by maximising the distance from the nearest data points of each class to the given hyperplane. When the data are not linearly separable, kernel transformations are used to map the data points onto a higher dimensional space where they can be separated by hyperplanes. 
    
    We use radial basis functio (RBF) kernels, also known as Gaussian kernels, which tend to cluster close points (with respect to the Euclidean distance) to the same class. This is based on the fact that we consider that the distribution of similarity features is directly related to the SED and therefore the same types of stars will have comparable similarity features.

    \section{Simulated data}
    \label{sect:sims}

    To simulate synthetic star observations we use the WaveDiff\footnote{\href{https://github.com/CosmoStat/wf-psf}{https://github.com/CosmoStat/wf-psf}} PSF model. WaveDiff provides a wavefront-based parametric PSF simulator that addresses the spectral and spatial dependence of the PSF across the FOV. This makes it possible to evaluate the PSF model at any desired wavelength, for any arbitrary position in the FOV. See \autoref{apx:wavediff} for details on WaveDiff PSF modelling.
    With the WaveDiff PSF simulator we can simulate observations of distant stars at any position in the FOV. To do so, we approximate the integral in \autoref{eq:obs_star} by \autoref{eq:sim_star}.
    We note that the SED of the star and the simulated PSF of the telescope are discretised in this approximation. In the following paragraphs we provide further details regarding the PSF model and the selection of SEDs and stellar types.
    
    \subsection{WaveDiff PSF simulator}
    
    \begin{figure}
        \subfloat[]{
        \includegraphics[clip,width=.99\columnwidth]{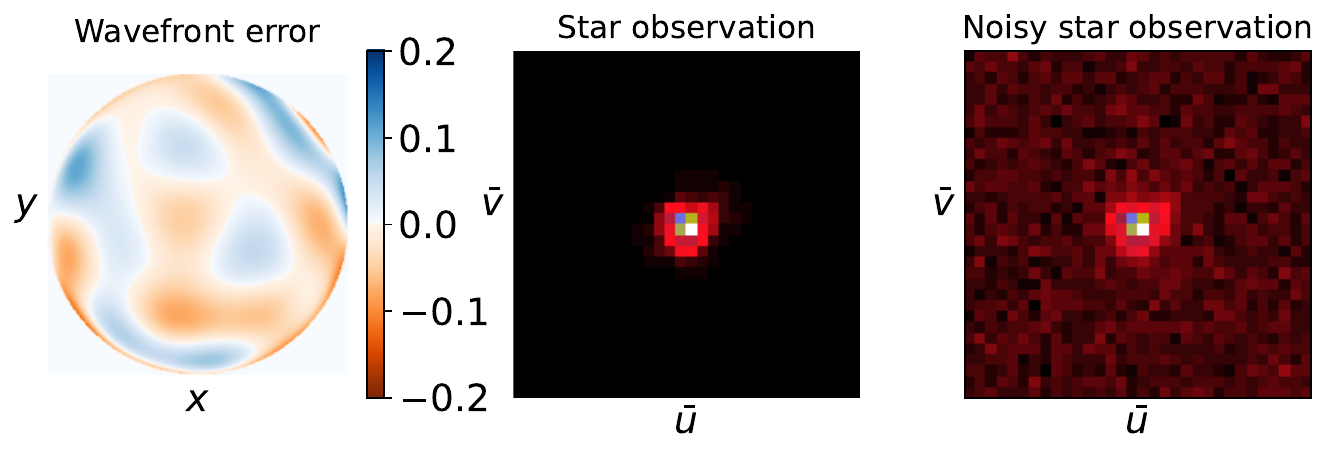}}
        
        \subfloat[]{   
        \includegraphics[clip,width=.99\columnwidth]{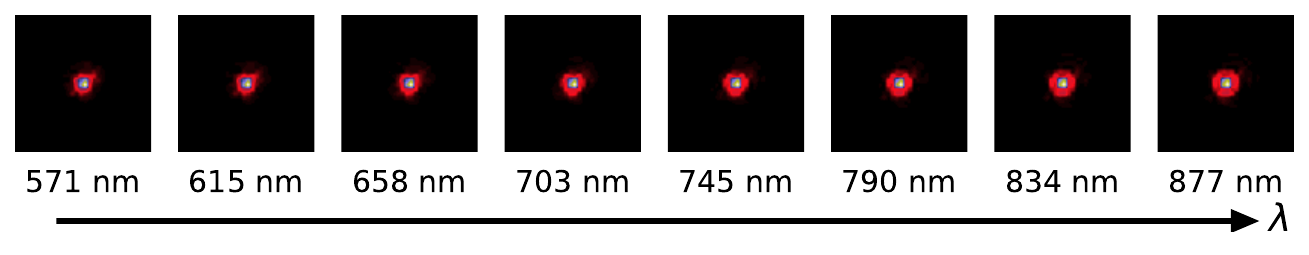}}
        \caption{WaveDiff simulations. (a) Wavefront error, noiseless star observation, and noisy star observation at a particular position in the FOV for a simulated PSF field. (b) Monochromatic PSFs for multiple wavelength values.}
        \label{fig:wfe_psf_example}
    \end{figure}
    
    \Autoref{fig:wfe_psf_example} shows different components of the PSF simulator: the WFE, the monochromatic PSFs, and the star observations (with and without noise). 
    First, the WFE for a random position in the FOV is shown in the first column of the top panel.
    Then, eight monochromatic PSFs computed from the aforementioned WFE are shown in the bottom panel. 
    Finally, the middle and right-hand side columns of the top panel show an example of a star observation with and without added noise. 

    The noise is modelled pixel-wise by an additive independent Gaussian random variable of zero mean and standard deviation $\sigma_{S/N}$. The total amount of noise added to a simulated observation depends on the desired signal-to-noise ratio $S/N$ (defined as \citet{liaudat}) as follows,
    \begin{equation}
            \sigma^2_{S/N} = \frac{\|I_{\text{star}}(\bar{u}, \bar{v}|u_i,v_i)\|^2_{\bar{uv}}}{S/N \;N_{pix}^2},
    \end{equation}
    where $N_{pix}$ is the stamp size (width and height) of the simulated star observations and $\|\cdot\|^2_{\bar{uv}}$ is the Frobenius matrix squared norm as defined in \autoref{eq:frobenius}.
    
    \subsection{SED templates}
    
\begin{figure}
        \centering
        \includegraphics[width=\columnwidth]{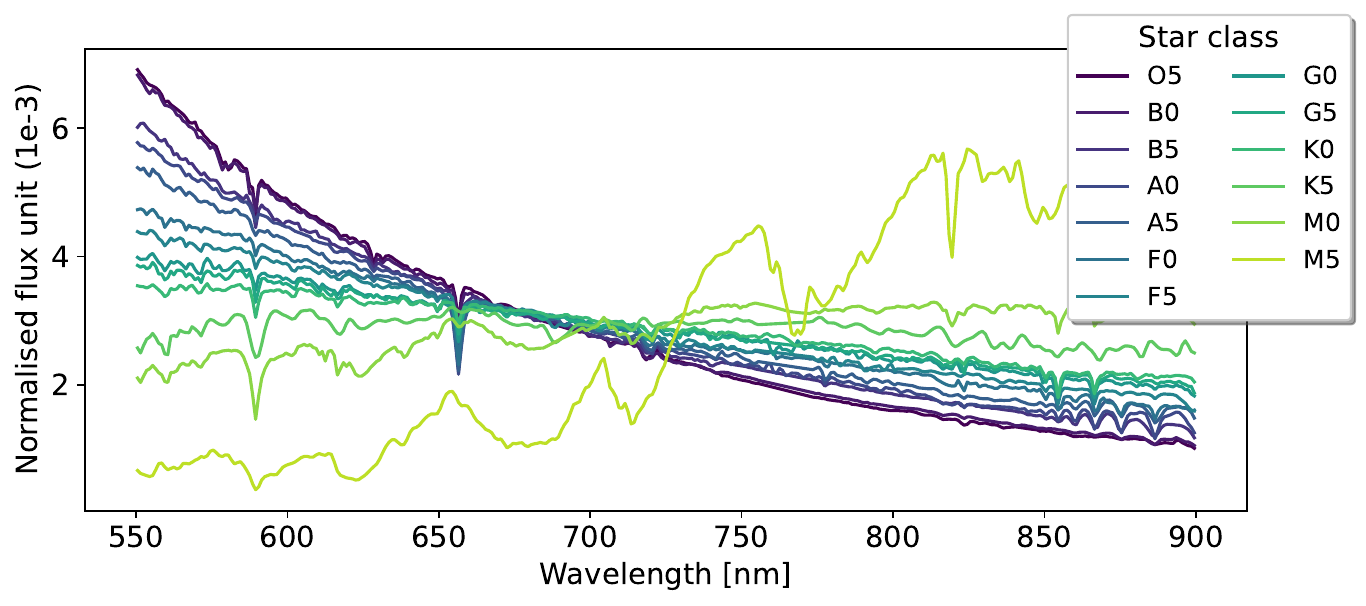}
        \caption{Spectral templates for the 13 stellar classes taken from \citet{Pickles_1998}. Spectra are limited to a \textit{Euclid}-like passband $[550 - 900]$ nm, with a resolution of $1$ nm. Spectra are flux normalised to unit sum.}
        \label{fig:pickles}
    \end{figure}

    The PSF simulator allows us to obtain monochromatic PSFs from the simulated WFE at any position in the FOV. To simulate a stellar observation, we sum the monochromatic PSFs over the wavelength weighted by the SED of the star as described by \autoref{eq:sim_star}. 
    Therefore, we need the spectral information of the stars. For this, we use 13 SED templates from \citet{Pickles_1998} corresponding to the following star types: $O5$, $B0$, $B5$, $A0$, $A5$, $F0$, $F5$, $G0$, $G5$, $K0$, $K5$, $M0$, and $M5$. The spectra are limited to the passband of the \textit{Euclid} VIS instrument \citep{euclidcollaboration2024euclidiivisinstrument}, from $550$ to $900$ nm, to simulate \textit{Euclid}-like star observations. In \autoref{fig:pickles}, we present the flux-normalised spectrum template $f_{\text{star}}(\lambda)$ for each stellar type. To obtain the observed star simulation as described in \autoref{eq:sim_star}, we compute the discrete SED of the star by integrating the spectrum over $n_\lambda$ regular wavelength bins $b^k$, matching the number of monochromatic PSFs. The centre of each bin corresponds to the wavelength of each monochromatic PSF. The bins, of size $\Delta b^k$ are computed as follows,
    \begin{equation}
        \text{SED}_{b^k}(\lambda_k) = \frac{1}{w_{b^k}} \int_{\lambda_k-\Delta b^k/2}^{\lambda_k+\Delta b^k/2} f_{\text{star}}(\lambda) d\lambda,
    \end{equation}
    where $w_{b^k}$ normalises the bin such that $\sum_{k=1}^{n_\lambda}\text{SED}_{b^k}(\lambda_k) = 1$.
    \subsection{Simulation parameters}
    The simulations used in this work are built from a random realisation of a WaveDiff PSF field, mainly governed by the following parameters: 
    $n_Z$, the maximum Zernike order in the WFE representation; 
    $d_{\text{max}}$, the degree of the polynomial variation of the Zernike coefficients $C_k(x,y)$ across the FOV; 
    $n_\lambda$, the number of spectral bins, which corresponds to the number of monochromatic PSFs and SED bins; 
    and $S/N$, the signal-to-noise-ratio range for the stellar observations. 
    The selection of these parameters depends mainly on the telescope that is taken as a reference for simulating the PSF and corresponding observations. The trade-off between the closeness of the simulations to real observations, and the available memory resources and computing power is also taken into consideration. We consider for the WFE a maximum Zernike order $d_Z = 45$ and a polynomial spatial variation of its coefficients of degree $d_{\text{max}} = 4$. 
    The number of spectral bins to be used is limited by the computational resources and the resolution of the available SEDs. For each wavelength value of the SED, the monochromatic PSF must be computed. Therefore, a larger number of bins requires a linear increase in time and memory resources. However, a larger number of bins allows for more realistic simulations, thereby making \autoref{eq:sim_star} a closer approximation to \autoref{eq:obs_star}. Simulated observations are generated with eight spectral bins to speed up the computation time for both the generation of the observations and the training of the PSF models. 
    For the results shown in the following sections, we consider that eight bins are sufficient to capture the spectral information of the star in the single-band simulated observation. Furthermore, \autoref{fig:similarity} demonstrates a clear visual separation of spectral classes based on the computed similarity features, further supporting that eight spectral bins provide sufficient resolution for our simulations.
    Finally, we vary the additive noise level for each simulated star so that the signal-to-noise ratio falls in the range $[20-110]$, which corresponds to the standard deviation of the Gaussian additive noise $\sigma$ falling approximately in the range $(10^{-3};\;2\times10^{-3})$. The pre-noise observations produced by WaveDiff are flux-normalised to one (i.e. the sum of the pixels is equal to one). 
    \Autoref{tab:parameters} summarises the values selected for the simulation parameters, as well as other relevant features of the simulations. The parameters concerning the dimensions and optics of the telescope and its characteristics, such as focal length, aperture radius, obscurations, passband, etc., are set as in \citet{liaudat} considering a \textit{Euclid}-like telescope.

    \begin{table}
        \centering
        \caption{WaveDiff simulation parameters.}
        \label{tab:parameters}
        \begin{tabular}{|l|l|l|}
        \hline
        \rowcolor[HTML]{E1E1E1} 
        Parameter                      & Description                & Value           \\ \hline
        $n_Z$                          & Maximum Zernike order      & $45$            \\ \hline
        $d_{\text{max}}$               &$C_k(x,y)$ polynomial degree& $4$             \\ \hline
        $n_\lambda$                     & Number of spectral bins    & $8$             \\ \hline
        $N_{\text{pix}}$               & Simulated observations size& $32$ px         \\ \hline
        $K$                            & Number of stellar classes  & $13$            \\ \hline
        $S/N$                          & Signal to noise ratio range& $[20-110]$      \\ \hline
        $\text{WFE}_{rms}$                      & Maximum WFE rms value      & $50$ nm         \\ \hline
        \end{tabular}
        \vspace{\baselineskip}
    \end{table}

    \section{Experiments}
    \label{sect:results}

    \begin{figure*}
        \centering
        \includegraphics[width=\linewidth]{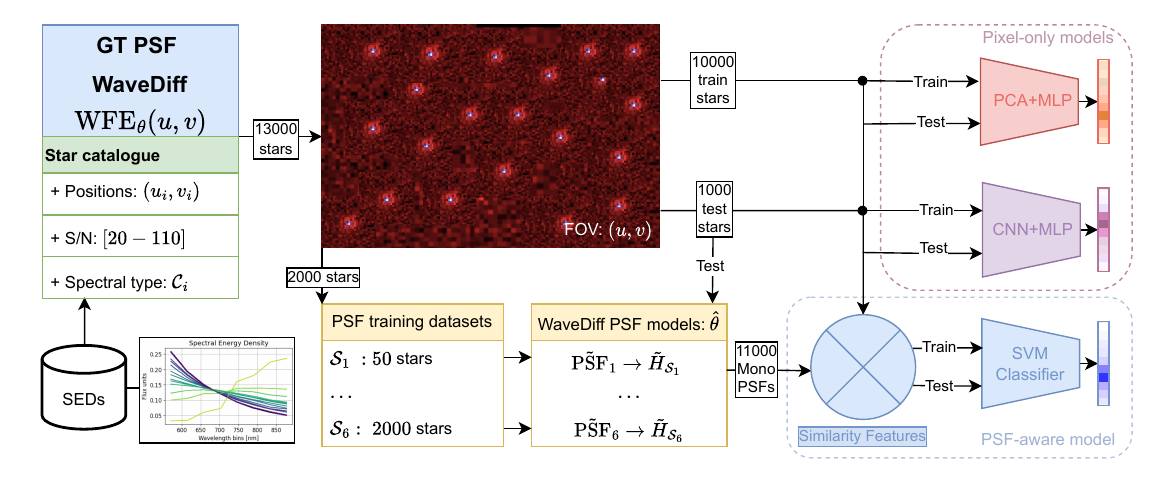}
        \caption{Project workflow: $13\,000$ star observations were simulated with the WaveDiff PSF simulator. Each star observation is located at a particular FOV position $(u_i,v_i)$, has a random noise level $S/N$ and belongs to a spectral class $\mathcal{C}_i$. $10\,000$ stars are directly used for training the pixel-only classification models, and $1\,000$ stars to test them. $2\,000$ stars are used to optimise the approximate PSF models required to calculate similarity features. The quality of the PSF models is evaluated with the $1\,000$ test stars. Once the train and test similarity features have been computed, the PSF-aware stellar classifier is trained with $10\,000$ stars and tested with $1\,000$ stars. }
        \label{fig:model_scenario}
    \end{figure*}

    Using the WaveDiff PSF simulator we generate a total of $13\,000$ star observations from a single ground truth PSF model. The star positions $(u_i,v_i)$ are selected randomly over the FOV with a uniform distribution. The associated SED for each simulated observation is chosen randomly from the $13$ spectral classes $\mathcal{C}_i$. The choice of a flat stellar type distribution is intended to mitigate any classification bias due to sample imbalance.
    \Autoref{fig:model_scenario} summarises all the models and simulation tools presented in this article.
    Star observation images are directly used to train ($10\,000$ stars) and test ($1\,000$ stars) the pixel-only classification models (PCA+MLP and CNN+MLP). The remaining $2\,000$ stars are divided into six nested datasets used to train six approximate PSF models, which are tested with the $1\,000$ test stars. We compute the monochromatic PSFs for every approximate PSF model at every star position in the FOV and we use them to compute the similarity features for the PSF-aware model. 
    
    In this section, we present the classification results for each method. First, we define the metrics used to compare the models. Then, we show the classification results of the pixel-only algorithms, validating \citet{kuntzer} results and contrasting them with our CNN approach. Finally, we present the performance of our novel PSF-aware classification method.
    
    \subsection{Classification metrics}
    To assess the performance of multi-class classification tasks, there exists a large variety of metrics, such as F-score, precision, recall, accuracy, cross entropy, etc. \citep{grandini2020metricsmulticlassclassificationoverview}. Many of these metrics are based on the confusion matrix (CM); we will therefore devote special attention to it.
    \subsubsection{Confusion matrix}
    The CM is a cross table that for each available class enumerates the number of assignments to the output classes. Each row of the table corresponds to the predicted labels $\hat{y}$ for a specific class $\mathcal{C}_i$ of the input data. The diagonal of the table, or matrix, shows the number of correctly classified elements per class ($\hat{y}=y$). Formally, the elements of the CM are defined as follows,
    \begin{equation}
        \text{CM}_{ij} = \sum_{x \in \mathcal{C}_i} \mathbbm{1}[\hat{y}=i],
    \end{equation}
    where $f(x)=\hat{y}$ is the predicted class for the element $x$.
    \subsubsection{Precision, recall, \& F1-score}
    Precision and recall are the main building blocks of binary classification metrics. In binary classification, precision is the fraction of correctly retrieved elements (true positives) and the total number of retrieved (claimed positives, that is true positives plus false positives) instances; and recall is the ratio between the true positives and the total number of relevant elements.
    \begin{equation}
        \begin{split}
            \text{Precision} = \frac{\text{TP}}{\text{TP}+\text{FP}},
        \end{split}
        \hspace{.2\columnwidth}
        \begin{split}
            \text{Recall} = \frac{\text{TP}}{\text{TP}+\text{FN}}.
        \end{split}
    \end{equation}
    The F1-score is widely used in binary classification because it leverages the number of correctly detected instances, true positives (TP) and true negatives (TN), and the number of missed instances, false positives (FP) and false negatives (FN). The F1-score is the harmonic mean of precision and recall,
    \begin{equation}
        \text{F1} = \frac{2}{\text{Precision}^{-1}+\text{Recall}^{-1}} = 
        \frac{2\text{TP}}{2\text{TP}+\text{FP}+\text{FN}}.
    \end{equation}
    The precision and recall, and thus the F1-score, can be generalised to multi-class classification by computing them class by class in an one-vs-all scenario, and then averaging over every class. 
    
    \subsubsection{Accuracy}
    Another relevant multi-class classification metric we consider is the accuracy, defined as the ratio between the number of correctly classified elements and the sample size. This metric is derived from the trace of the confusion matrix,
    \begin{equation}
        \text{Accuracy} = \frac{
        Tr\left( \text{CM} \right)}{
        \sum_{ij}\text{CM}_{ij}
        }.
    \end{equation}
    Given the proximity between neighbouring stellar types, which have a difference of half a spectral class, a top-two accuracy metric is considered. This is mainly motivated by the similarity between two spectra of neighbouring classes (see \autoref{fig:pickles}), which when discretised into $8$ wavelength bins the difference between two neighbouring spectral classes lies in the range of the SEDs photometric measurement noise. The top-two accuracy allows for neighbouring stellar type misclassification. These cases are located on the super diagonal and sub diagonal of the CM, defining the top-two accuracy metric as follows,
    \begin{equation}
        \text{Top-two accuracy} = \frac{
        \sum_{i=j+1}\text{CM}_{ij}  + 
        \sum_{i=j}\text{CM}_{ij} + 
        \sum_{i=j-1}\text{CM}_{ij} }{
        \sum_{ij}\text{CM}_{ij}
        }.
    \end{equation}
    In \citet{kuntzer}, this metric is referred to as the success rate.
    
    \subsection{Pixel-only classification}
    \begin{table}
    \begin{center}
    \caption{Classification metrics for the PCA+MLP, CNN+MLP, and SVM+PSF models. }
    \label{tab:results}
    \begin{tabular}{@{}lccc@{}}
    \toprule
    \midrule
    Model            & F1    & Accuracy & Top-two accuracy \\
    \midrule
    PCA+MLP          & 0.366 & 0.370    & 0.757            \\
    CNN+MLP          & 0.385 & 0.391    & 0.746            \\
    \midrule
    SVM+$\text{PSF}_{\mathcal{S}_1}$    & 0.392 & 0.410    & 0.755   \\
    SVM+PSF$_{\mathcal{S}_4}$ & 0.506 & 0.512 & 0.873 \\
    SVM+PSF$_{\text{GT}}$          & 0.546 & 0.549    & 0.910            \\
    \midrule
    \bottomrule
    \end{tabular}
    \end{center}

    Notes: $\text{SVM+PSF}_{\mathcal{S}_1}$ stands for the SVM+PSF classifier that uses the similarity features computed with an approximate PSF model trained on the $\mathcal{S}_1$ dataset, which has a relative error of $2.4\%$. Analogously for the $\mathcal{S}_4$ dataset with $500$ stars and a relative error of $1\%$. The SVM+PSF$_{\text{GT}}$ row uses the ground truth PSF to compute the similarity features.
    \end{table}
    We train, evaluate, and compare the pixel-only classification methods: $\text{PCA+MLP}$ and $\text{CNN+MLP}$.
    Both methods were trained with the same dataset of $10\,000$ simulated star observations. We evaluate both methods on the $1\,000$ test dataset using the aforementioned classification metrics. The results are presented in \autoref{tab:results}. Each row of the table corresponds to a different model, and the first two rows correspond to the pixel-only methods. The first column corresponds to the one-vs-all F1-score averaged over all the classes, the second column is the accuracy of each model, and the third one is the top-two accuracy. 
    
    We note that the metrics for the PCA+MLP method are consistent with those of \citet{kuntzer}. The performance of the model is slightly lower, but this is expected as the data we are using has a higher complexity in the spatial variation of the PSF, higher noise levels, and the total number of samples used for training is lower.

    While using a convolutional network instead of the PCA decomposition uses state-of-the-art deep learning techniques and gives more flexibility to the extraction of spatial features, we do not observe a significant improvement in classification. We believe this is primarily due to the degeneracy between PSF and spectral type, which imposes a limit on the accuracy of pixel-only classification, regardless of which technique is used. 
    
\begin{figure}
        \subfloat{
        \includegraphics[trim={0 1.6cm 0 0},clip,width=.99\columnwidth]{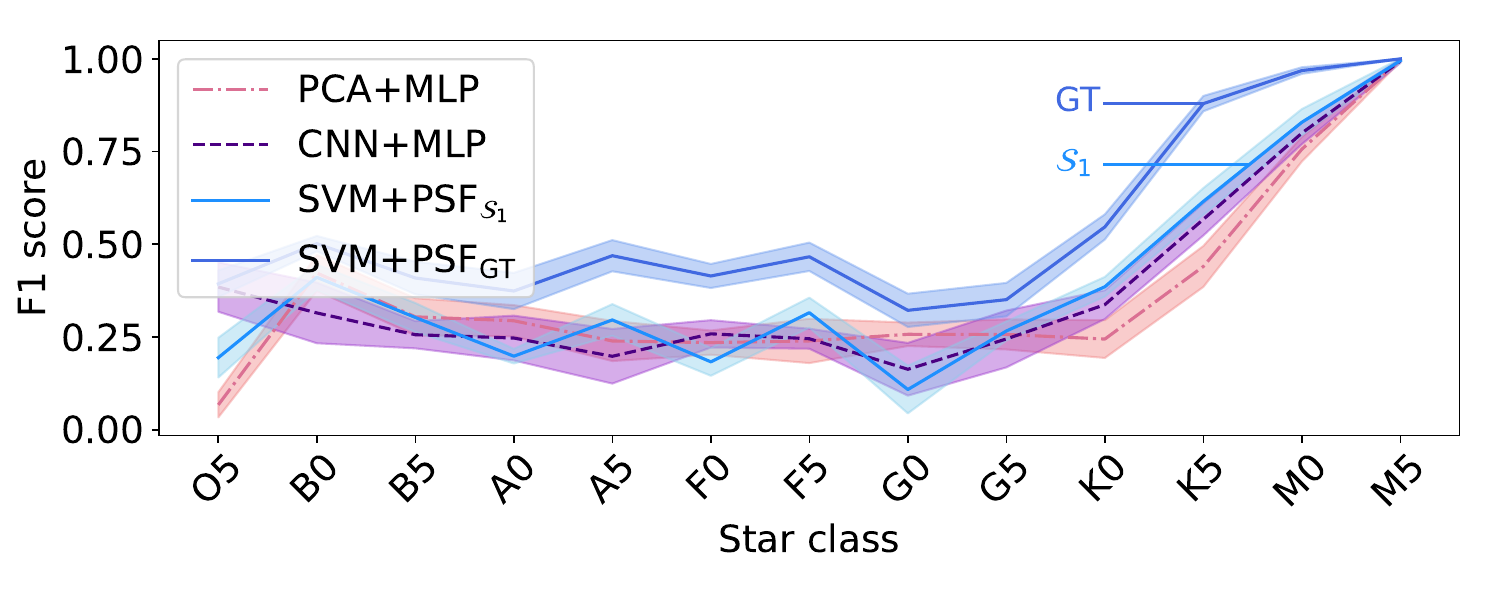}}
        
        \subfloat{   
        \includegraphics[trim={0 1.6cm 0 0},clip,width=.99\columnwidth]{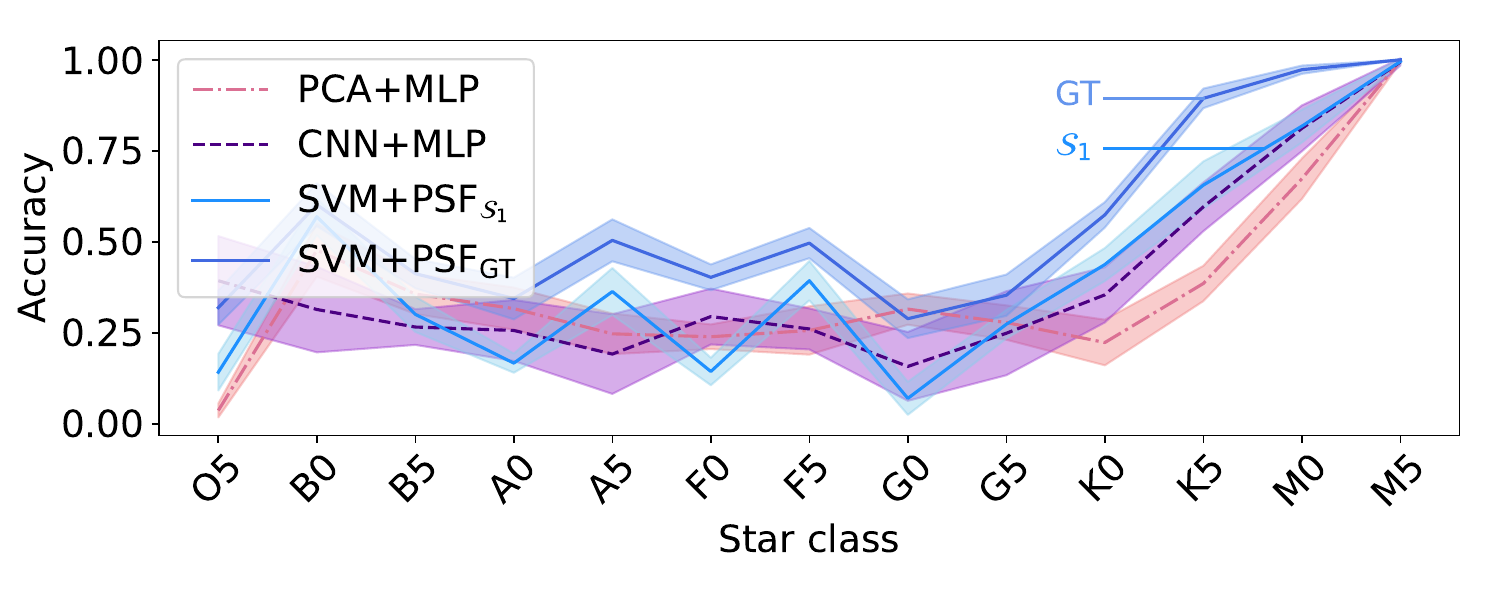}}
        
        \subfloat{   
        \includegraphics[clip,width=.99\columnwidth]{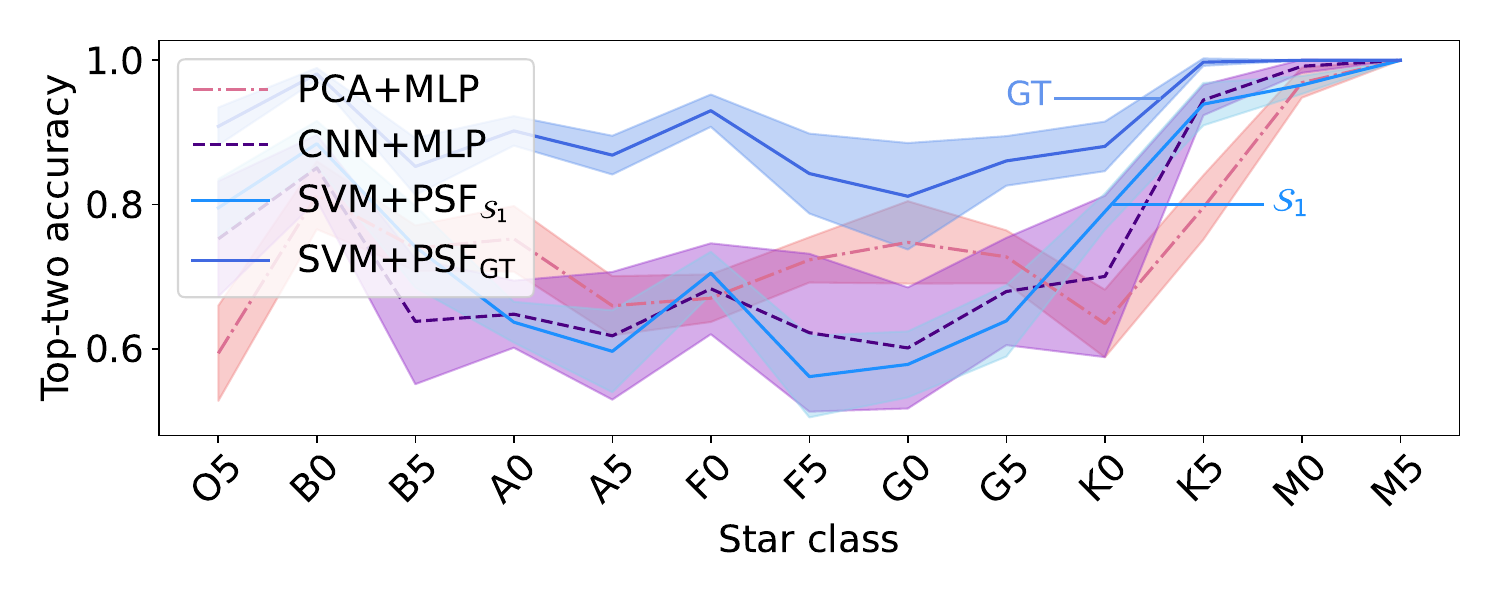}}
        
        \caption{F1 score, accuracy, and top-two accuracy by stellar type for the PCA+MLP, CNN+MLP, and SVM+PSF models. The $\text{SVM+PSF}_{\mathcal{S}_1}$ line shows the SVM+PSF classifier that uses the similarity features computed with an approximate PSF model trained on the $\mathcal{S}_1$ dataset, which has a relative error of $2.5\%$. }
        \label{fig:one_vs_all_metrics}
    \end{figure}

    In addition to the metrics shown in \autoref{tab:results}, which are averaged across all stellar classes, we show the one-vs-all metrics for each class in \autoref{fig:one_vs_all_metrics}. The exact values are detailed in \autoref{apx:one-v-all}. We observe that the one-vs-all F1-score accuracy and top-two accuracy have a similar distribution for both the PCA+MLP model (in red dot-dashed line) and the CNN+MLP model (in violet dashed line).
    We note that these metrics are higher for the redder stars. This is expected when examining the spectra shown in \autoref{fig:pickles}. The figure demonstrates that the spectral differences between adjacent stellar types are larger for red stars (M-type) compared to blue stars (O-type).
    This is also consistent with the distribution of similarity features shown in \autoref{fig:similarity}, where we see that the red stars (and neighbouring types) have a similarity feature distribution that is easily distinguishable by eye from the other types. 

    \begin{figure}
        \centering
        \hspace*{-.5cm}
        \includegraphics[width=.99\columnwidth, trim={0 0 0 0},clip]{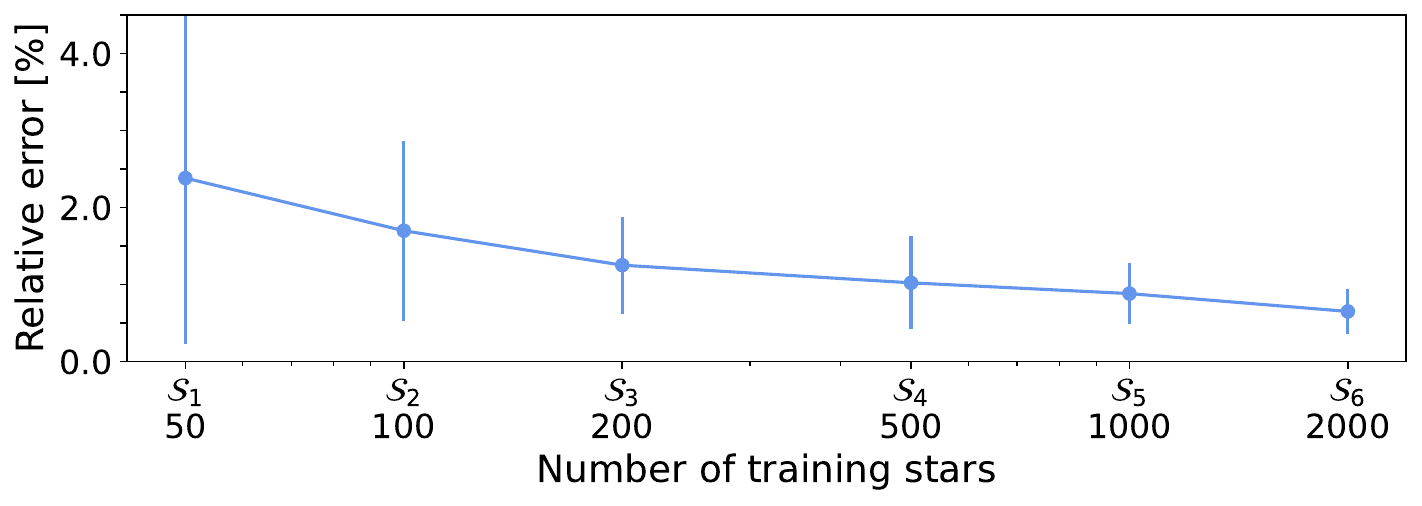}
        \caption{Relative error at observation resolution for each approximate PSF model as a function of the number of training stars. The error bars represent the standard deviation of the relative errors of the stars in the test dataset.}
        \label{fig:PSF_approx}
    \end{figure}

    \begin{figure}
        \centering
        \hspace*{-.5cm}
        \includegraphics[width=\columnwidth, trim={0cm 0 .5cm 0},clip]{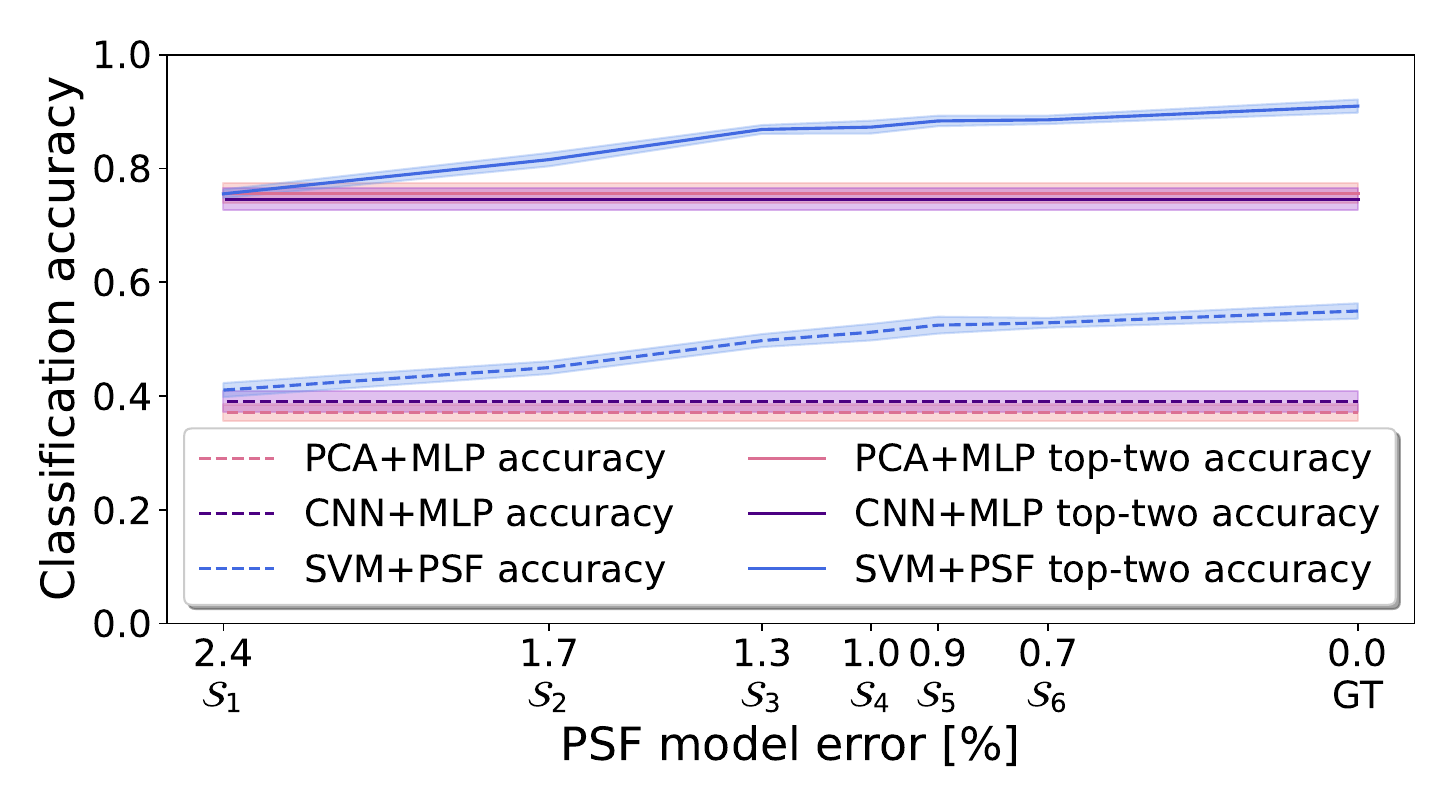}
        \caption{Accuracy (dashed blue line) and top-two accuracy (continuous blue line) metrics for the PSF-aware model (SVM+PSF) as a function of the PSF model error. In red and violet (horizontal lines), corresponding metrics for the pixel-only classification methods (PCA+MLP and CNN+MLP) are plotted as a reference.}
        \label{fig:results}
    \end{figure}

    \subsection{PSF-aware classification}
    Before training the PSF-aware stellar classifier, we compute the approximate PSF models with differently sized datasets. We present below the estimated PSF models and then the results of the PSF-aware stellar classifier.

    \subsubsection{Approximated PSF models}

    As mentioned in \autoref{sect:psf_modeling}, we need to produce an approximate model of the PSF to compute the similarity features of each star, as the ground truth PSF is in principle unknown. We use WaveDiff to compute an approximate PSF model from the star observations.
    We employ $2\,000$ simulated star observations for training the approximate PSF model. To assess various levels of PSF modelling uncertainty and analyse the impact on stellar SED classification, we subdivide the $2\,000$ stars into six datasets with increasing number of stars, where each dataset is contained in the next $\mathcal{S}_1 \subset \mathcal{S}_2 \subset ... \subset \mathcal{S}_6$. Since each star observation is a sample of the GT PSF, we expect that a larger number of stars will lead to a more accurate PSF model. We fit a WaveDiff PSF model $\tilde{H}(u,v;\lambda)$ to each of the datasets listed in \autoref{tb:datasets}. \Autoref{fig:PSF_approx} shows the relative error of each PSF model as a function of the number of training stars contained in the dataset. The relative error is defined as
    \begin{equation}
        Err_{\text{rel}} = \frac{1}{n}\sum_{i=1}^n \frac{\text{RMS}(I_{star}^{(i)}-\hat{I}_{star}^{(i)})}{\text{RMS}(I_{star}^{(i)})}\times100\%,
    \end{equation}
    where $I_{star}$ is the target PSF, $\hat{I}_{star}$ is the predicted PSF, $n$ is the total number of test stars and $\text{RMS}(\cdot)$ is the root mean square of a matrix as defined in \citet{liaudat}. For each PSF model in \autoref{fig:PSF_approx} the relative error is averaged over the $1\,000$ test stars. These six approximate PSF models provide the basis for computing the similarity features used in the classification algorithm.
    \begin{table}
    \centering
    \caption{List of nested datasets with the corresponding number of stars.}
    \label{tb:datasets}
    \begin{tabular}{|l|c|c|c|c|c|c|}
    \hline
    \rowcolor[HTML]{E1E1E1} 
    Dataset & $\mathcal{S}_1$ & $\mathcal{S}_2$   & $\mathcal{S}_3$   & $\mathcal{S}_4$   & $\mathcal{S}_5$    & $\mathcal{S}_6$    \\ \hline
    Size    & 50 & 100 & 200 & 500 & 1\,000 & 2\,000 \\ \hline
    \end{tabular}
    \end{table}
    The resulting PSF relative error at observation resolution\footnote{In this study, we evaluate the PSF only at observation resolution and not at super-resolution as is done in \cite{liaudat}.} is around $2.5\%$ for dataset $\mathcal{S}_1$ containing $50$ stars, and $0.7\%$ for dataset $\mathcal{S}_6$ containing $2\,000$ stars. The relative error range of the PSF models is consistent with what is shown in Fig. 9 of \citet{liaudat} therefore we assume that we cover a sufficient range of PSF model accuracy for our analysis.
    
    \subsubsection{Classification results}
    With each approximated PSF model we evaluate the eight monochromatic PSFs, $\tilde{H}(\bar{u}, \bar{v}; \lambda_k|u_i,v_i)$, at each star position $(u_i,v_i)$ in the classification dataset. From the monochromatic PSFs and the star observations we compute the similarity features following \autoref{eq:similarity}.
    We fit the the proposed SVM algorithm to the similarity features, independently for each dataset $\mathcal{S}$, and make the predictions for each star in the test dataset. We also consider the scenario in which we have perfect knowledge of the PSF by using the ground truth PSF for computing the similarity features. This sets a baseline for the best possible performance of the PSF-aware classification method.
    
    The classification accuracy and top-two accuracy are presented in  \autoref{fig:results}. The detailed F1-score and accuracy values are provided in \autoref{tab:SVM_PSF_results}. The PSF-aware method, SVM+PSF, outperforms both pixel-only classifiers (PCA+MLP and CNN+MLP) for every considered error level of the PSF. We obtain a $91\%$ top-two accuracy with perfect knowledge of the PSF, and $76\%$ top-two accuracy with the least precise PSF model we tested. We recall that this model was trained with only $50$ stars in the FOV, which is much less than what we expect for a \textit{Euclid}-like survey \citep{laureijs}. In addition, we note that the approximate PSF model trained with the $2\,000$ fiducial stars allows for a $89\%$ top-two classification accuracy, which is very close to that obtained with the ground truth PSF model. Also, models trained with fewer stars (between $200$ and $2\,000$) lead to a similar classification performance. These approximate PSF models, although not accurate enough for weak-lensing analyses \citep{10.1093/mnras/stt384, 10.1093/mnras/sts371, PSFcal}, are able to assist stellar classification methods by breaking the degeneracy between PSF size and spectral type, and outperforming pixel-only classification methods.

    \subsubsection{Improving the final PSF model}
    
    To demonstrate a potential application of our novel PSF-aware classifier we test how much of an impact additional stars with classified SEDs in a given FOV would have on improving the final PSF model. The aim is to extract more information from the observations and improve the quality of the PSF model.
    
\begin{figure}
        \includegraphics[trim={.6cm .5cm .7cm .5cm},clip,width=\columnwidth]{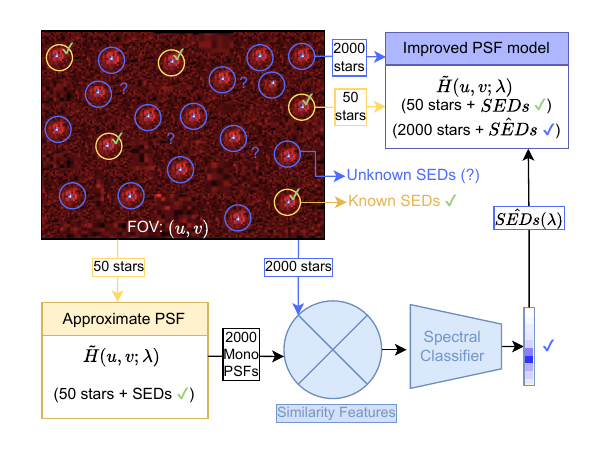}
        \caption{PSF improvement scenario. The observed exposure contains 2\,050 stars, of which 50 have complementary spectral information (in yellow) and 2000 do not (in blue). An approximate PSF model is trained from 50 stars with known SEDs. The approximate PSF is used in the context of the PSF-aware classifier to spectrally classify the remaining 2\,000 stars. The classified stars are assigned a SED template and can be used to train an improved PSF model.}
        \label{fig:psf_improvement_scenario}
    \end{figure}

    \Autoref{fig:psf_improvement_scenario} shows a diagram of the proposed PSF improvement scenario. We assume a single Euclid-like wide-band exposure. A fraction of the unresolved stars present in this exposure have known SEDs from complementary measurements (in yellow) and the remaining do not (in blue). 
    We propose to obtain an approximate PSF model from the stars with known SEDs. Then, to use this approximate PSF model with our PSF-aware classifier to assign SEDs to the remaining stars. Finally, we can attempt to improve the PSF model considering both the stars with measured SEDs and the ones with assigned SED templates. We expect the final PSF model to have a lower error as it uses more samples of the underlying PSF and the spatial distribution of the additional samples better captures the spatial variation of the PSF in the FOV.

\begin{figure}
        \includegraphics[trim={.0cm .0cm .0cm .0cm},clip,width=\columnwidth]{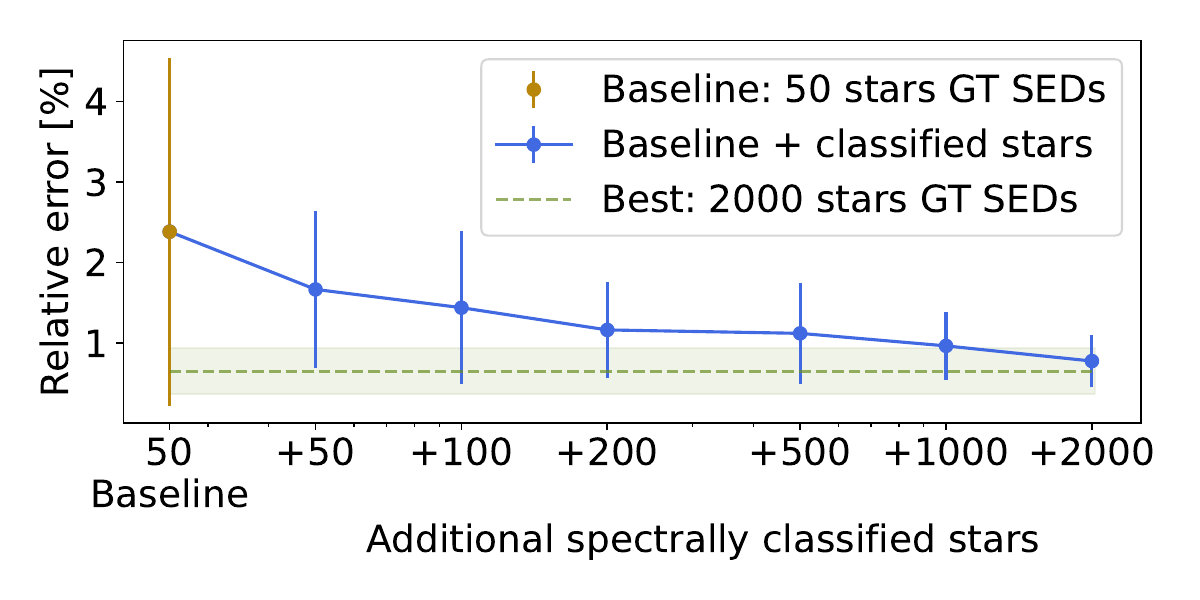}
        \caption{Relative PSF error, at observation resolution, as a function of the number of training stars. The baseline is set to the performance of the approximate PSF trained with 50 stars and ground truth SEDs (dark yellow). The subsequent PSF models (blue points) are trained using the 50 baseline stars in addition to increasing numbers of stars with SEDs assigned by our PSF-aware classifier. The error bars represent the standard deviation of the relative errors of the test dataset. The green dashed line shows the idealised minimum relative PSF error that can be obtained when using 2\,000 stars with ground truth SEDs, and the green shaded area represents the standard deviation of the relative errors.}
        \label{fig:psf_improvement_results}
    \end{figure}

    As a proof of concept, we test the proposed PSF improvement scenario using a sample of 2\,050 simulated star observations.
    We train WaveDiff using 50 stars with GT SEDs and obtain an approximate PSF model with a relative error of $2.4\%$ at observation resolution. We then use the approximate PSF in our PSF-aware classifier to assign SED templates to the remaining 2\,000 stars. For this purpose we employ an SVM classifier pre-trained on 8\,000 similarity feature samples. We obtain a classification accuracy of $41\%$ and a top-two accuracy of $76\%$, which is consistent with the results presented in \autoref{sect:results}. Finally, the 2\,000 newly classified stars are used together with the original 50 stars to train new WaveDiff PSF models. We study how the PSF error varies as we increase the total number of training stars. \Autoref{fig:psf_improvement_results} shows the relative error, at observation resolution, of the PSF model as a function of the number of training stars. In dark yellow, we present the baseline relative error of the approximate PSF model (i.e. trained on the 50 stars with GT SEDs). The relative error of subsequent PSF models that use the additional sample of spectrally classified stars is plotted in blue. We show that, as we increase the number of stars with classified SEDs, the relative error of the PSF decreases. 
    Using the full sample of 2\,000 spectrally classified stars, we achieve a relative error of $0.78\%$. This represents a PSF error reduction of almost $70\%$. Finally, the dotted line (in green) represents the relative error obtained by training the PSF model with 2\,000 stars and the corresponding GT SEDs (i.e. an idealised performance assuming unlimited spectroscopic counterparts) for comparison. The minimum error is $0.65\%$, only slightly lower than that obtained using the sample of stars with classified SEDs. 

    We emphasise that this is a highly idealised and simplified test case and significant work would still need to be carried out to conclusively demonstrate the applicability of this approach to real survey data. However, the initial results are promising and indicate that it may be possible to improve PSF modelling performance in single wide-band images by increasing the star sample with classified SED stars.

    \section{Conclusions}
    \label{sect:conclusion}
    The SED of observed stars is crucial for chromatic PSF modelling of wide-field single-band telescopes. However, SED measurements are often expensive and scarce for low-brightness stars, limiting the number of stars available for PSF modelling. A reliable spectral classification method using survey data could significantly benefit PSF modelling. Assigning SED templates to observed stars would increase the number of stars available for constraining the PSF model. This approach could enhance the accuracy of PSF modelling and, consequently, improve various astronomical studies, particularly weak gravitational lensing analysis.
    
    In this paper, we propose a novel method for spectral classification from single wide-band observations of stars.
    This new method, referred to as the PSF-aware classifier, incorporates the spectral variation of an approximate PSF model of the telescope in order to break the degeneracy between the size of the PSF and the spectral type of stars. To evaluate the performance of our PSF-aware method, we compare it with pixel-only classifiers that rely solely on the star image pixel values.
    We implement and validate the results of the pixel-only classifier presented in \cite{kuntzer} and propose an update based on a convolutional neural network. We find that the CNN method performs about the same as the \cite{kuntzer} approach. We emphasise that these classifiers, since they use only the pixel values of the observed stars, do not address the degeneracy between the PSF size and the spectral type of the star. Consequently, we introduce the PSF-aware stellar classification method and address how the PSF modelling error impacts the spectral classification accuracy.
    
    We show how the PSF-aware model breaks the aforementioned degeneracy, pushing the classification accuracy further and outperforming both pixel-only classification methods by around $10\%$. We obtain a top-two accuracy of $91\%$ with the proposed model and perfect knowledge of the PSF. We also study how the level of fidelity of the PSF model impacts the classification metrics, resulting in a top-two accuracy of $87\%$ with a PSF model trained with $500$ stars ($1\%$ relative error over the low-resolution PSF samples), and a top-two accuracy of $76\%$ with a PSF model trained with only $50$ stars ($2.4\%$ relative error over the low-resolution PSF samples). This shows that the approximate PSF models, which assist the classifier, although not sufficiently precise for WL analyses, are useful for breaking the degeneracy and improving the classification accuracy. We use WaveDiff \citep{liaudat} to obtain approximate PSF models for the purposes of the work presented. However, we expect similar performance from any PSF modelling method, provided it can model the spectral variation of the PSF.

    We then test the PSF-aware classifier in a proof-of-concept study, where we evaluate how much the additional stars with classified SEDs in a given FOV improve the modelling of the final PSF. We show that PSF models trained with complementary classified stars allow the relative PSF error to be reduced by almost $70\%$. The inclusion of spectrally classified stars reduces the relative error of the PSF from $2.5\%$, for an approximate reference model trained with 50 stars and their GT SEDs, to an error of $0.78\%$ when using 2\,000 complementary stars spectrally classified with our PSF-aware classifier. While this experiment is rather simplistic and does not fully represent the complexity of PSF modelling from real data, the results obtained are promising and illustrate a potential use of the proposed spectral classifier.  

    Future studies can explore improvements to the PSF-aware classifier, such as replacing the SVM classifier with a neural network or using convolutional networks 
    to compute custom features optimised for the stellar classification problem. The next steps for the work presented would include making more realistic simulations by adding redshift information to the star observations and increasing the number of wavelength bins used for stellar observations generation (\autoref{eq:sim_star}). In this case, we would need to study the selection of the number of similarity features, which we set equal to the number of spectral bins, in more detail. By addressing these issues, we would move significantly closer to applying our PSF-aware classifier to real survey data. This could be of interest for space missions such as \textit{Euclid}, where the spectral information (SEDs) of the observed stars is crucial for training the PSF model. 

    \begin{acknowledgements}
         This work was supported by the TITAN ERA Chair project (contract no. 101086741) within the Horizon Europe Framework Program of the European Commission, and the Agence Nationale de la Recherche (ANR-22-CE31-0014-01 TOSCA).
         This work is additionally supported by the European Community through the grant ARGOS (contract no. 101094354).
         This work was granted access to the HPC resources of IDRIS under the allocation 2023-AD011012983R2 made by GENCI.
    \end{acknowledgements}

%
%

\bibliographystyle{aa}
\bibliography{astro}

\begin{appendix} 
\onecolumn

\section{PSF modelling notation}
\label{apx:notation}

\begin{table}[h]
    \centering
    \caption{Coordinates and notation used throughout this article. (\cite{Liaudat_2023}).}
    \label{tab:notation}
        \begin{tabular}{cl} 
        \toprule
        \midrule
        Variable & \multicolumn{1}{c}{Description}   \\
        \midrule
        \multicolumn{2}{c}{\textit{Coordinates}} \\
        \midrule
        $(x,y)$                 & Pupil plane or output aperture plane coordinates. \\
        $(u,v)$                 & Image or focal plane coordinates. \\
        $(\xi, \eta)$           & Object plane coordinates. \\
        $(\bar{u}, \bar{v})$    & Pixel coordinates, the discrete counterpart of the image plane.  \\
        $\mathbf{p}_i$          & 3D spatial coordinate. \\
        $\lambda$               & Wavelength  \\
        $t$                     & Time  \\
        \midrule
        \multicolumn{2}{c}{\textit{Notation}} \\
        \midrule
        $\mathcal{I}, \mathcal{H}, \ldots$  & Calligraphic uppercase variables are continuous functions. \\
        $I, H, \ldots$                      & Uppercase variables are matrices. \\
        $c_m, b_{1}^{k}, \ldots$            & Lowercase variables are scalars. \\
        $I_\mathrm{img}(\bar{u},\bar{v};t|u_i,v_i) \in \mathbb{R}$ & Pixel value at position $(\bar{u},\bar{v})$ for the image $I_\mathrm{img}$ with its  centroid at position $(u_i,v_i)$ \\
        & observed at time $t$. \\
        $I_{\mathrm{img}, (\cdot|u_i,v_i)} \in \mathbb{R}^{p \times p}$ & Observed image  with its centroid at position $(u_i,v_i)$.  \\
        \midrule
        \bottomrule
        \end{tabular}
    
\end{table}

\section{PSF modelling}
    
    PSF modelling \citep{Liaudat_2023} can be divided into two main categories: parametric models and data-driven models. 
    Parametric models rely on building a physical representation of the optical system.
    After the physical model is built there is a reduced set of parameters to adjust, but it can be very challenging to tune them properly, requiring costly measurements of the instrument, precise calibrations, and time-consuming optical simulations.
    Data-driven or non-parametric models do not necessarily represent the physics of the instrument and have a higher degree of freedom, which allows optimal parameter combinations to be found using optimisation techniques. Such models typically use observations to constrain the PSF representation.

    This approach is based on the observational model for an image centred at the position $(u_i,v_i)$ in the FOV,
    \begin{equation}
        I_{\text{img}}(\bar{u},\bar{v}|u_i,v_i) = 
        \mathcal{F}_p \left\{ \int_0^{+\infty} \mathcal{T}(\lambda)\; 
        (\mathcal{I}_{\text{GT}} \star \mathcal{H}_{\text{int}})
        (u,v;\lambda | u_i, v_i) \; d\lambda \right\} 
        \;+\; N(\bar{u},\bar{v}|u_i,v_i).
    \end{equation}
    The ground truth image $\mathcal{I}_{\text{GT}}$ is convolved with the PSF of the telescope $\mathcal{H}_{\text{int}}$ (intensity impulse response) and integrated over the passband of the telescope, which is given by the transmission function  $\mathcal{T}(\lambda)$. The $\mathcal{F}_p$ operator is a discretisation function that models the pixelisation of the detector (sampling) and $N$ embodies the observational noise. The observed image $I_{\text{img}}$, with pixel coordinates $(\bar{u}, \bar{v})$, is a discrete version of the ground truth image corrupted by the PSF of the telescope and the observational noise.

    Observations of distant stars are particularly interesting for PSF modelling. Unresolved stars can be considered as point sources represented by a 2-dimensional delta distribution with a chromatic dependence described by the spectral energy distribution of the star,
    \begin{equation}
        \mathcal{I}_{\text{star}}(u,v;\lambda|u_i,v_i) = \text{SED}(\lambda)\;\delta(u-u_i,v-v_i).
    \end{equation}

    The convolution between the delta function and the PSF will result in a sample of the PSF at the $(u_i, v_i)$ position. Therefore each observation of an unresolved star provides a sample of the PSF of the telescope at the corresponding position in the FOV \citep{annurev:/content/journals/10.1146/annurev-astro-081817-051928,10.1093/mnras/staa3679}. 
    The observational model of a distant star is as follows
    \begin{equation}
        I_{\text{star}}(\bar{u},\bar{v}|u_i,v_i) = 
        \mathcal{F}_p \left\{ \int_0^{+\infty} \mathcal{T}(\lambda) \text{SED}(\lambda) \; \mathcal{H}_{\text{int}}(u,v;\lambda|u_i, v_i) \; d\lambda \right\} 
        \;+\; N(\bar{u},\bar{v}|u_i,v_i),
    \end{equation}

    \noindent where $\mathcal{H}_{\text{int}}(u,v;\lambda|u_i, v_i)$ is the PSF of the telescope with its centre at the position of the star. The PSF has two spatial coordinates $(u,v)$ and one spectral coordinate $\lambda$. The PSF sample is integrated together with the star SED over the passband of the telescope given by the transmission function $\mathcal{T}(\lambda)$. The $\mathcal{F}_p$ operator is a discretisation function that models the pixelisation of the detector (sampling) and $N$ embodies the observational noise. The observed image $I_{\text{star}}$, with pixel coordinates $(\bar{u}, \bar{v})$, is a single-band discrete version of the star corrupted by the PSF of the telescope and the observational noise.

    \section{WaveDiff}
    \label{apx:wavediff}
    WaveDiff \citep{liaudat} is a data-driven PSF model that operates in the wavefront space and is based on a differentiable optical module that models the physics of the optical process to go from the wavefront error (WFE) to the pixel-level PSF. By simplifying the optical system of the telescope to a single converging lens, the Fraunhofer diffraction approximation can be applied, thus relating the PSF at the focal plane and the wavefront error at the pupil plane as follows  \citep[Eq. 5]{liaudat},
    \begin{equation}
        H(\bar{u}, \bar{v};\lambda|u_i, v_i) \propto
        \left| \text{FT}
        \left\{ 
            \exp\left[
            \frac{2 \pi i}{\lambda} \text{WFE}(x, y| u_i, v_i)
            \right]  
        \right\}
        \left[\frac{\bar{u}}{\lambda f_L},\; \frac{\bar{v}}{\lambda f_L}\right]
        \right|^2
        ,
    \label{eq:optical_syst}
    \end{equation}
    
    \noindent where FT is the Fourier transform, $u$ and $v$ are the focal plane coordinates, $x$ and $y$ are the pupil plane coordinates and $f_L$ is the focal length of the optical system.

    The WaveDiff WFE parametric representation is composed of a weighted sum of Zernike polynomials \citep{Noll:76}, which are widely used by the optics community to represent the phase of spherical wavefronts. Zernike polynomials are two-dimensional functions that are orthogonal to each other over the unit disc. The Zernike amplitudes vary across the FOV coordinates $(u,v)$ for each Zernike order $l$ and are represented by a two-dimensional function $C_l^P(u,v)$. In addition, the WFE has a non-parametric contribution, whose features, $S^{NP}_m(x,y)$, are trained together with the spatial variation coefficients, $C^{NP}_m(u,v)$, using unresolved stars observations. The WFE as a function of the FOV coordinates $(u,v)$ can be modelled as follows,
    \begin{equation}
        WFE(x,y|u, v) = \overbrace{\sum_{l=1}^{n_Z}C^P_l(u,v|\theta)\; Z_l(x,y)}^{parametric} \;\;+ 
        \underbrace{\sum_{m=1}^{n_{NP}}C^{NP}_m(u,v|\theta)\; S^{NP}_m(x,y|\theta)}_{non-parametric},
    \end{equation}
    where $Z_l$ is the $l$-th order Zernike polynomial and $n_Z$ is the maximum order considered for the WFE representation. In this model, the spatial variation of the Zernike coefficients is a polynomial of degree $d_{\text{max}}$.
    
    WaveDiff employs single-band star observations to optimise the parameters $\theta$, modelling both the spatial and spectral variations of the GT PSF.

\clearpage

\section{One-vs-all metrics}
\label{apx:one-v-all}

\begin{table}[h]
    \centering
    \caption{F1-score for each classification model, for each individual class.}
    \label{tb:f1_score}
        \begin{tabular}{@{}lccccccccccccc@{}}
        \toprule
        \midrule
        \multirow{2}{*}{Model} & \multicolumn{13}{c}{Stellar type}                                                        \\ \cmidrule(l){2-14} 
                               & O5   & B0   & B5   & A0   & A5   & F0   & F5   & G0   & G5   & K0   & K5   & M0   & M5   \\
        \midrule
        PCA+MLP                & 0.07 & 0.42 & 0.3  & 0.29 & 0.24 & 0.23 & 0.24 & 0.26 & 0.26 & 0.24 & 0.44 & 0.76 & 1.0  \\
        CNN+MLP                & 0.38 & 0.31 & 0.26 & 0.25 & 0.2  & 0.26 & 0.25 & 0.16 & 0.24 & 0.34 & 0.57 & 0.8  & 0.99 \\
        SVM+$\text{PSF}_{\mathcal{S}_1}$        & 0.19 & 0.41 & 0.3  & 0.2  & 0.3  & 0.18 & 0.32 & 0.11 & 0.27 & 0.39 & 0.62 & 0.83 & 1.0  \\
        SVM+PSF                & 0.39 & 0.5  & 0.41 & 0.37 & 0.47 & 0.41 & 0.47 & 0.32 & 0.35 & 0.55 & 0.88 & 0.97 & 1.0  \\
        \midrule
        \bottomrule
        \end{tabular}
\end{table}

\begin{table}[h]
    \centering
    \caption{Accuracy for each classification model, for each individual class.}
    \label{tb:accuracy_per_class}
    \begin{tabular}{@{}lccccccccccccc@{}}
    \toprule
    \midrule
    \multirow{2}{*}{Model} & \multicolumn{12}{c}{Stellar type}                                                      \\ \cmidrule(l){2-14} 
                           & O5 & B0        & B5   & A0   & A5   & F0   & F5   & G0   & G5   & K0   & K5   & M0   & M5   \\ \midrule
    PCA+MLP                   & 0.04 & 0.48 & 0.36 & 0.32 & 0.25 & 0.24 & 0.26 & 0.31 & 0.28 & 0.22 & 0.39 & 0.67 & 1.0  \\
    CNN+MLP                   & 0.39 & 0.31 & 0.27 & 0.26 & 0.19 & 0.29 & 0.26 & 0.16 & 0.25 & 0.35 & 0.6  & 0.81 & 0.99 \\
    SVM+$\text{PSF}_{\mathcal{S}_1}$                & 0.14 & 0.57 & 0.3  & 0.17 & 0.36 & 0.14 & 0.39 & 0.07 & 0.27 & 0.44 & 0.66 & 0.82 & 1.0  \\
    SVM+PSF                   & 0.32 & 0.6  & 0.41 & 0.34 & 0.5  & 0.4  & 0.5  & 0.29 & 0.35 & 0.57 & 0.89 & 0.97 & 1.0  \\ 
    \midrule
    \bottomrule
    \end{tabular}
\end{table}

\begin{table}[h]
    \centering
    \caption{Top-two accuracy for each classification model, for each individual class.}
    \label{tb:top-two_per_class}
    \begin{tabular}{@{}lccccccccccccc@{}}
    \toprule
    \midrule
    \multirow{2}{*}{Model} & \multicolumn{13}{c}{Stellar type}                                                       \\ \cmidrule(l){2-14} 
                           & O5   & B0   & B5   & A0   & A5   & F0   & F5   & G0   & G5   & K0   & K5   & M0   & M5  \\ \midrule
    PCA+MLP                    & 0.59 & 0.81 & 0.74 & 0.75 & 0.66 & 0.67 & 0.72 & 0.75 & 0.73 & 0.63 & 0.8  & 0.97 & 1.0 \\
    CNN+MLP                   & 0.75 & 0.85 & 0.64 & 0.65 & 0.62 & 0.68 & 0.62 & 0.6  & 0.68 & 0.7  & 0.94 & 0.99 & 1.0 \\
    SVM+$\text{PSF}_{\mathcal{S}_1}$                & 0.8  & 0.88 & 0.74 & 0.64 & 0.6  & 0.7  & 0.56 & 0.58 & 0.64 & 0.79 & 0.94 & 0.97 & 1.0 \\
    SVM+PSF                   & 0.91 & 0.98 & 0.85 & 0.9  & 0.87 & 0.93 & 0.84 & 0.81 & 0.86 & 0.88 & 1.0  & 1.0  & 1.0 \\ 
    \midrule
    \bottomrule
    \end{tabular}
\end{table}

\section{SVM+PSF results}
\label{apx:resutls}
\begin{table}[h]
    \begin{center}
    \caption{Classification results for the SVM+PSF classifier.}
    \label{tab:SVM_PSF_results}
    \begin{tabular}{@{}lccc@{}}
    \toprule
    \midrule
    Model & \multicolumn{1}{c}{F1} & \multicolumn{1}{c}{Accuracy} & \multicolumn{1}{c}{Top-two accuracy} \\ \midrule
    PCA+MLP          & 0.366 & 0.370    & 0.757\\
    CNN+MLP          & 0.385 & 0.391    & 0.746\\
    \midrule
    SVM+PSF$_{\mathcal{S}_1}$ & 0.392 & 0.410 & 0.755 \\
    SVM+PSF$_{\mathcal{S}_2}$ & 0.431 & 0.450 & 0.816 \\
    SVM+PSF$_{\mathcal{S}_3}$ & 0.491 & 0.498 & 0.869 \\
    SVM+PSF$_{\mathcal{S}_4}$ & 0.506 & 0.512 & 0.873 \\
    SVM+PSF$_{\mathcal{S}_5}$ & 0.516 & 0.525 & 0.884 \\
    SVM+PSF$_{\mathcal{S}_6}$ & 0.523 & 0.529 & 0.886 \\
    SVM+PSF$_{\text{GT}}$ & 0.546 & 0.549 & 0.910 \\ 
    \midrule
    \bottomrule
    \end{tabular}
    \end{center}

    Notes: Each row shows a different PSF model used for computing the similarity features. The pixel-only classification metrics correspond to the first two rows.
\end{table}
    
\label{LastPage}

\clearpage
\twocolumn
\end{appendix}

\end{document}